\documentclass[useAMS,usenatbib]{mn2e}

\usepackage{graphicx}
\usepackage{mathptmx}
\usepackage[hang,raggedright]{subfigure}
\graphicspath{{./Images/}}

\begin{document}
\title[Radio source stacking]{Radio source stacking and the infrared /
  radio correlation at microJy flux densities}

\author[T.\ Garn \& P.\ Alexander]{Timothy Garn$^{1,2}$\thanks{E-mail:
    tsg@roe.ac.uk} and Paul Alexander$^{2}$\\ 
    $^{1}$SUPA, Institute for Astronomy, Royal Observatory Edinburgh,
    Blackford Hill, Edinburgh EH9~3HJ\\
    $^{2}$Astrophysics Group, Cavendish Laboratory, 19 J.~J.~Thomson
    Ave., Cambridge CB3~0HE}

\date{\today}
\pagerange{\pageref{firstpage}--\pageref{lastpage}; } \pubyear{2007}
\pubyear{2008}
\label{firstpage}
\volume{000}
\maketitle

\begin{abstract}
We investigate the infrared / radio correlation using the technique of
source stacking, in order to probe the average properties of radio
sources that are too faint to be detected individually.  We compare
the two methods used in the literature to stack sources, and
demonstrate that the creation of stacked images leads to a loss of
information.  We stack infrared sources in the {\it Spitzer}
extragalactic First Look Survey (xFLS) field, and the three northern
{\it Spitzer} Wide-area Infrared Extragalactic survey (SWIRE) fields,
using radio surveys created at 610~MHz and 1.4~GHz, and find a
variation in the absolute strength of the correlation between the xFLS
and SWIRE regions, but no evidence for significant evolution in the
correlation over the 24-$\umu$m flux density range 150~$\umu$Jy --
2~mJy.  We carry out the first radio source stacking experiment using
70-$\umu$m-selected galaxies, and find no evidence for significant
evolution over the 70-$\umu$m flux density range 10~mJy -- 100~mJy.
\end{abstract}


\section{Introduction}
There is a well-known relationship between the infrared and radio flux
densities of star-forming galaxies \citep*[the `infrared / radio
correlation', e.g.][]{Helou85,Appleton04} that is thought to be the
result of both the infrared and radio emission from star-forming
galaxies being related to the rate of massive star formation.  Recent
studies of the infrared / radio correlation within nearby galaxies
\citep[e.g.][]{Murphy06} give weight to this argument, as the radio
emission has been shown to represent a smeared out version of the
infrared emission.  This has been attributed to the diffusion of
cosmic ray electrons away from star-forming regions.  The correlation
is seen to apply to a wide range of galaxies associated with
star-formation, but not to those sources which are associated with
Active Galactic Nuclei (AGN) activity
\citep[e.g.][]{deJong85,Sanders88,Sopp91,Roy98}.

Studies at high redshift are tentatively finding that the infrared /
radio correlation does not deviate significantly from that seen in the
local Universe, at least out to $z \sim 1$
(\citealp[e.g.][]{Garrett02,Appleton04,Frayer06}), and potentially to
$z \sim 3.5$ \citep{Ibar08}.  The correlation has been shown to remain
invariant over more than 5 orders of magnitude \citep*[e.g.][]{Yun01},
however studies that rely on samples of sources which are detected at
both infrared and radio wavelengths will naturally be mainly sampling
luminous galaxies, and there has been relatively little work carried
out on testing the infrared / radio correlation within fainter
galaxies.

The technique of source `stacking' is well established as a method for
combining data from many individual objects in order to study the
statistical properties of sources which would otherwise be below the
detection limit for a particular survey.  By constructing a list of
source positions based on prior information obtained at another
frequency, a series of small `cut-out' images can be created, centred
on these source positions, of sources that may be below the noise
level (and therefore undetected).  If $N$ of these images are
combined, each with a local noise of $\sigma$, then the noise level of
the stacked image will be expected to decrease approximately as
$\sigma / \sqrt{N}$, which quickly allows sources below the original
detection threshold of the survey to be studied.  Stacking has
previously been carried out using optical \citep[e.g.][]{Zibetti05}
and infrared \citep[e.g.][]{Zheng06,Zheng07} images, as well as in the
radio.  The large coverage area of the Faint Images of the Radio Sky
at Twenty-cm \citep*[FIRST;][]{Becker95} survey has led to a
number of stacking experiments being carried out
\citep[e.g.][]{Wals05,deVries07,White07,Hodge08}, looking at the radio
properties of quasars and LL-AGN, and some studies of star-forming
galaxies have taken place using smaller, but deeper radio surveys
\citep[e.g.][]{Boyle07,Ivison07,Beswick08,Carilli08}.

\renewcommand{\thefootnote}{\it{\alph{footnote}}}
\begin{table*}
  \begin{center}
\begin{minipage}{\textwidth}
\begin{center}
  \caption{A summary of the properties of the radio surveys used in
  this work.  The area and resolution of each survey, and the noise
  level at the centre of each image are given, along with the primary
  reference paper, which should be consulted for further details.}
  \label{tab:surveydetails}
    \begin{tabular}{lccccccl}
\hline
Field & Instrument & Frequency & Area & Resolution & Position Angle & Noise level &
 Reference\\
 & & & (deg$^{2}$) & (arcsec$^{2}$) & (deg) & ($\umu$Jy~beam$^{-1}$) & \\
\hline
xFLS         & VLA  & 1.4~GHz & 4 & $5.0\times5.0$ & 0  & 23 & \citet{Condon03}\\
xFLS         & GMRT & 610~MHz & 4 & $5.8\times4.7$ & 60 & 30 & \citet{Garn07}\\
ELAIS-N1     & GMRT & 610~MHz & 9 & $6.0\times5.0$ & 45 &
40 / 70\footnote[1]{Deep / shallow regions respectively.} & \citet{Garn08EN1}\\
Lockman Hole & GMRT & 610~MHz & 5 & $6.0\times5.0$ & 45 & 60 & \citet{Garn08LH}\\
ELAIS-N2     & GMRT & 610~MHz & 6 & $6.5\times5.0$ & 70 & 90 & \citet{Garn09EN2}\\
\hline
    \end{tabular}
  \end{center}
\end{minipage}
\end{center}
\end{table*}
\renewcommand{\thefootnote}{\arabic{footnote}}

There have been two stacking studies of the infrared / radio
correlation for galaxies which are detected in the infrared, but are
below the detection limits of their radio surveys.  The strength of
the infrared / radio correlation can be quantified by the logarithmic
flux density ratio $q_{24}$ \citep[][and see
Section~\ref{sec:VLAstackq}]{Appleton04}, and the results from these
studies appear to be inconsistent, with \citet{Boyle07} finding a
significantly higher value of $q_{24}$ than is observed in sources
which are detected in both wavebands, and \citet{Beswick08} finding a
significantly lower value of $q_{24}$, and a tentative evolution in
the value of $q_{24}$ with 24-$\umu$m flux density.  In this work we
describe a stacking study of the infrared / radio correlation, using
the sensitive {\it Spitzer Space Telescope} \citep{Werner04}
observations of the {\it Spitzer} extragalactic First Look Survey
field (xFLS) and the {\it Spitzer} Wide-area Infrared Extragalactic
survey \citep[SWIRE;][]{Lonsdale03} fields.  The three northern SWIRE
fields are the European Large-Area {\it ISO} Survey-North~1
(ELAIS-N1), -North~2 (ELAIS-N2) and Lockman Hole regions, and we take
radio data from a 1.4-GHz Very Large Array (VLA) survey of the xFLS
field \citep{Condon03}, and 610-MHz Giant Metrewave Radio Telescope
(GMRT) surveys of the xFLS \citep{Garn07}, ELAIS-N1 \citep{Garn08EN1},
Lockman Hole \citep{Garn08LH} and ELAIS-N2 \citep[][in
prep.]{Garn09EN2} regions.

In Section~\ref{sec:techniques} we describe the two stacking
techniques which are used in the literature, and calculate the values
of $q_{24}$, using 1.4-GHz data from within the xFLS field.  In
Section~\ref{sec:stack610} we extend this work to the four 610-MHz
survey fields, and demonstrate that AGN contamination is not a
significant problem for our infrared samples.  We compare the results
from our four fields in Section~\ref{sec:stackingdiscussion}, and find
a field-to-field variation in the infrared / radio correlation, which
is too great to be explained through cosmic variance.  We compare our
results to those of \citet{Boyle07} and \citet{Beswick08}, and discuss
potential explanations for the systematic variation.  

\section{Stacking techniques}
\label{sec:techniques}
Previous stacking studies of the infrared / radio correlation have
been carried out at 1.4~GHz, and in order to compare our results to
those in the literature we initially stack infrared sources within the
xFLS field using the 1.4-GHz image of the region \citep{Condon03}.  A
summary of the characteristics of this survey, and the other radio
surveys used in this work, is given in Table~\ref{tab:surveydetails}.
Positional information for the infrared sources was taken from the
24-$\umu$m source catalogue of \citet{Fadda06}, which contains 16,905
extragalactic point sources above a signal-to-noise (S/N) of 5.  The
source catalogue is 50~per~cent complete over the field at a
24-$\umu$m flux density of 300~$\umu$Jy, with a flux density limit of
210~$\umu$Jy, while a deeper verification region, covering
approximately 10~per~cent of the field, is 50~per~cent complete to
150~$\umu$Jy, with a flux density limit of 120~$\umu$Jy.

Of the 16,905 sources in the catalogue, 14,820 are located within the
1.4-GHz image.  A small ($\sim0.1$~arcsec) positional offset between
the 24~$\umu$m catalogue and the radio image was identified and removed
before performing any stacking.  The xFLS 24-$\umu$m catalogue does not
classify sources by type, however we will demonstrate in
Section~\ref{sec:stacktype} that there is no significant contamination
in the catalogue from stars or AGN.  There are two principal methods
used in the literature to carry out stacking experiments, which are
described in Sections~\ref{sec:aperture} and \ref{sec:stackedimages}.

\subsection{Measuring the flux density of individual sources}
\label{sec:aperture}
The first method for carrying out source stacking measures the radio
flux density at each of the individual source positions, bins the
sources by their infrared flux density, and uses the statistical
distribution of radio flux density for sources within an infrared flux
density bin to calculate the typical source properties.  While the
individual measurements will be very uncertain, particularly for
sources near to, or below the noise level, the statistical properties
of the distribution should be robust \citep[e.g.][]{White07}.

The resolution of the 1.4-GHz xFLS image is $5\times5$~arcsec$^{2}$.
In order to calibrate the flux density measurements, a circular
aperture with radius of 4~arcsec was centred on the location of 79
moderately bright radio sources, and the total flux density within the
aperture measured.  An aperture correction factor was then calculated,
in order to match the measured flux densities of the bright sources to
their catalogued values \citep{Condon03}.  Any errors in flux density
measurement will propagate through the remainder of the analysis, so
the sources that were used for this calibration were required to be:
\begin{enumerate}
\item Brighter than 1~mJy total flux density, so that sources were
  detected with high significance and the catalogued flux density
  measurements were known to be accurate.
\item Fainter than 10~mJy total flux density.  This requirement is
  very important when calibrating the aperture size for the GMRT
  images in later sections, since $>$~10~mJy sources can be strongly
  affected by phase errors \citep[e.g.][]{Garn08EN1}.
\item Within the central region of the mosaic, so that any edge
  effects and increasing noise levels due to the varying primary beam
  would not affect the flux density calibration.
\item Unresolved, with a ratio of total to peak flux density $<$ 1.5.
  The faint star-forming galaxies being stacked are expected to be
  unresolved, with a typical size of $\sim1.5$~arcsec
  \citep{Beswick08} and a stacking analysis would not be appropriate
  for extended sources due to the different flux density distribution
  for each object.
\end{enumerate}
The calculated correction factor for a 4-arcsec aperture was 1.37 for
the VLA xFLS image (with similar values found for the four GMRT images
which are used later in this work).  This correction factor has been
calculated only from sources with flux densities between 1 and 10~mJy,
and may not be appropriate for fainter radio sources -- if this is the
case, then varying the aperture size (and hence the correction factor)
will lead to different results.  We tested aperture sizes between 2
and 10~arcsec -- after applying the correction factor, the flux
density measurements for the faint sources were consistent with each
other for aperture sizes between 2 and 5~arcsec.  At greater sizes,
the radio flux density calculated for the faint sources began to
increase.  We used a 4~arcsec radius aperture for all stacking
measurements described in this work, and all measurements made within
the aperture have had the correction factor applied unless otherwise
stated.

The flux density measurement technique described in this section sums
all of the radio flux density within a circular aperture, irrespective
of its origin.  There is the possibility that a 4~arcsec radius
aperture is sufficiently large that source confusion may be a problem
(where the radio emission from more than one source is incorrectly
assumed to all result from a single 24-$\umu$m target).  In order to
test for source confusion, we used the {\it Spitzer} Infrared Array
Camera \citep[IRAC;][]{Fazio04} observations of the xFLS field to
identify nearby counterparts to the 24-$\umu$m sources.  The IRAC
source catalogue \citep{Lacy05} contains 103,193 sources with
detections in at least one of the 3.6-, 4.5-, 5.8- and 8-$\umu$m bands.

We selected a sample of 9062 24-$\umu$m sources which were located
within the centre of the IRAC observations, and measured the number of
IRAC counterparts within 4~arcsec of each 24-$\umu$m source.  While
some of the 24-$\umu$m sources do not have an IRAC counterpart (917;
10.1~per~cent), the majority of the sources had a single IRAC
counterpart within 4~arcsec (7712; 85.1~per~cent), with only 409
sources (4.5~per~cent) having two counterparts and 24 sources
(0.3~per~cent) having three.  We conclude that confusion from nearby
faint sources will not significantly affect the radio flux density
measurements.

\begin{figure}
  \centerline{\subfigure[Measured and median binned radio flux density
    for the 14,820 24-$\umu$m sources within the xFLS 1.4-GHz image of
    \citet{Condon03}.  Note that the majority of sources are below the
    limit at which they could be detected individually, as shown by
    the upper dotted line, and would not be included in a conventional
    analysis.]{ \includegraphics[width=0.45\textwidth]{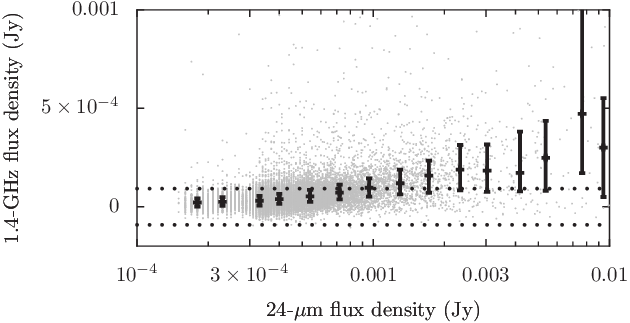}}}
    \centerline{\subfigure[Measured and median binned radio flux
    density for 14,820 random source positions, with each position
    assigned a 24-$\umu$m flux density from the catalogue at random.
    The radio flux density in each bin is consistent with zero.]{
    \includegraphics[width=0.45\textwidth]{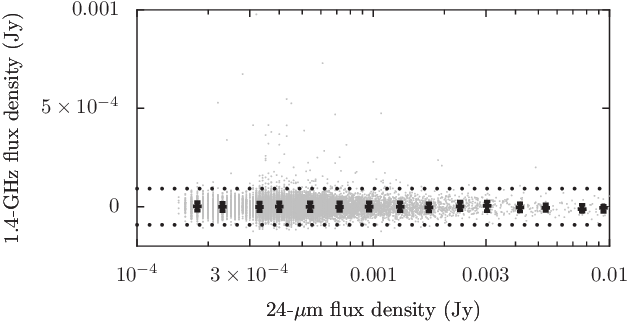}}}
  \caption{The 1.4-GHz radio flux density for each source in the
  24-$\umu$m catalogue, measured as described in
  Section~\ref{sec:aperture} (grey dots), the values of $\pm4\sigma$
  (horizontal black dotted lines) and the median values of flux
  density within each bin (black points with errors).  The error bars
  denote the size of the IQR of radio flux density in each bin.  The
  vertical striping seen in this and later figures is an artefact of
  the quoted precision of the 24-$\umu$m catalogue.  The decrease in
  the number of sources seen below a 24-$\umu$m flux density of
  300~$\umu$Jy is due to the smaller coverage area of the verification
  region -- see text for more details.}
  \label{fig:vlaflux}
\end{figure}

Fig.~\ref{fig:vlaflux}(a) shows the measured (aperture-corrected)
1.4-GHz radio flux densities for each source in the 24-$\umu$m
catalogue, along with $\pm4\sigma$ (where $4\sigma$ is the detection
limit for individual radio sources).  The figure is plotted on a
log-linear scale in order to show the range of radio flux density that
is measured, which can be negative due to the effects of noise.  The
decreased number of sources in Fig.~\ref{fig:vlaflux} (and later
figures displaying xFLS data) which have a 24-$\umu$m flux density of
$<300$~$\umu$Jy is due to the smaller coverage area of the deep
verification region of the survey.

In order to look at the statistical properties of sources, we bin them
logarithmically by their 24-$\umu$m flux densities, and calculate the
median infrared and radio flux density in each bin.  The median
estimator is more robust to outliers than the mean, and we will
demonstrate that the median is the most appropriate choice for this
data in Section~\ref{sec:stackedimages}.  We plot the median binned
flux densities in Fig.~\ref{fig:vlaflux}(a), with distribution widths
which denote the inter-quartile range (IQR) for each bin, a more
robust measurement of the range than the standard deviation.

For comparison, Fig.~\ref{fig:vlaflux}(b) shows the measured radio
flux densities from 14,820 random source positions in the image -- a
few locations have a measured flux density above the $4\sigma$
detection limit and appear to have a source present, but the remainder
of the random positions have measured flux density below the detection
threshold.  The median radio flux density for all bins is consistent
with zero.

\subsection{Creating stacked images}
\label{sec:stackedimages}
\begin{figure*}
  \centerline{\subfigure[150 -- 201~$\umu$Jy.]{
                \includegraphics[width=0.12\textwidth]{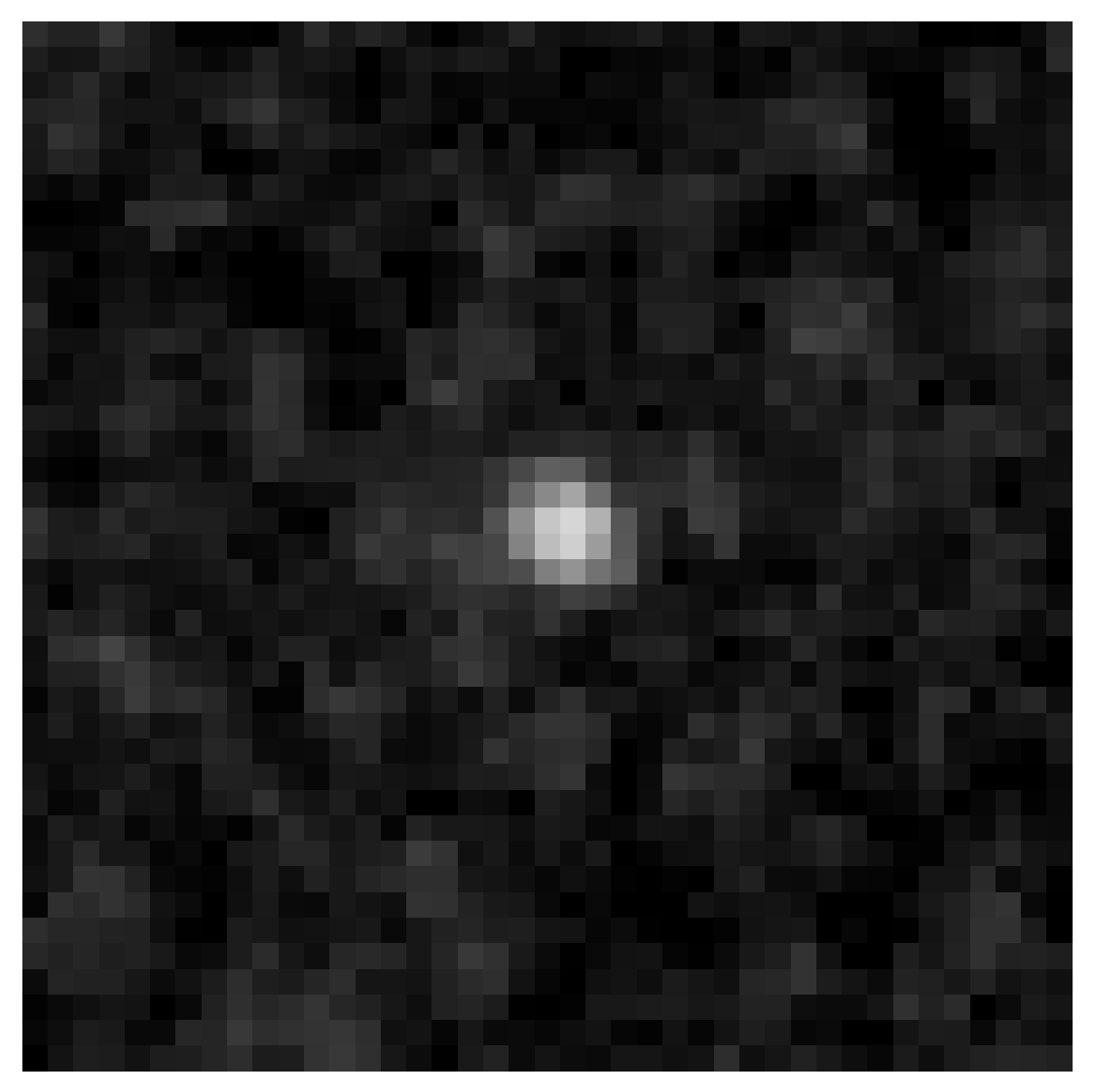}}
                \subfigure[201 -- 268~$\umu$Jy.]{
                \includegraphics[width=0.12\textwidth]{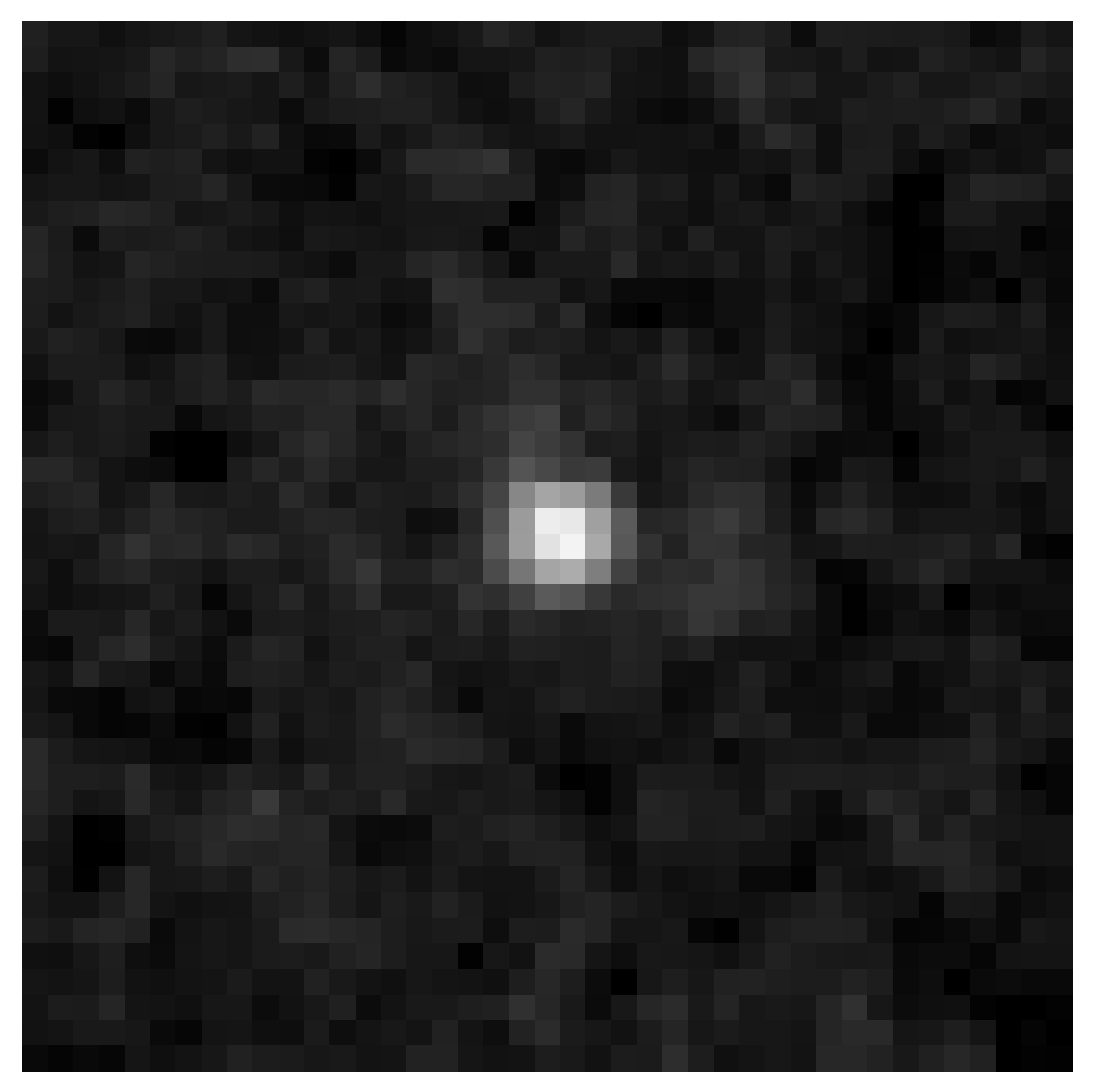}}
                \subfigure[268 -- 359~$\umu$Jy.]{
                \includegraphics[width=0.12\textwidth]{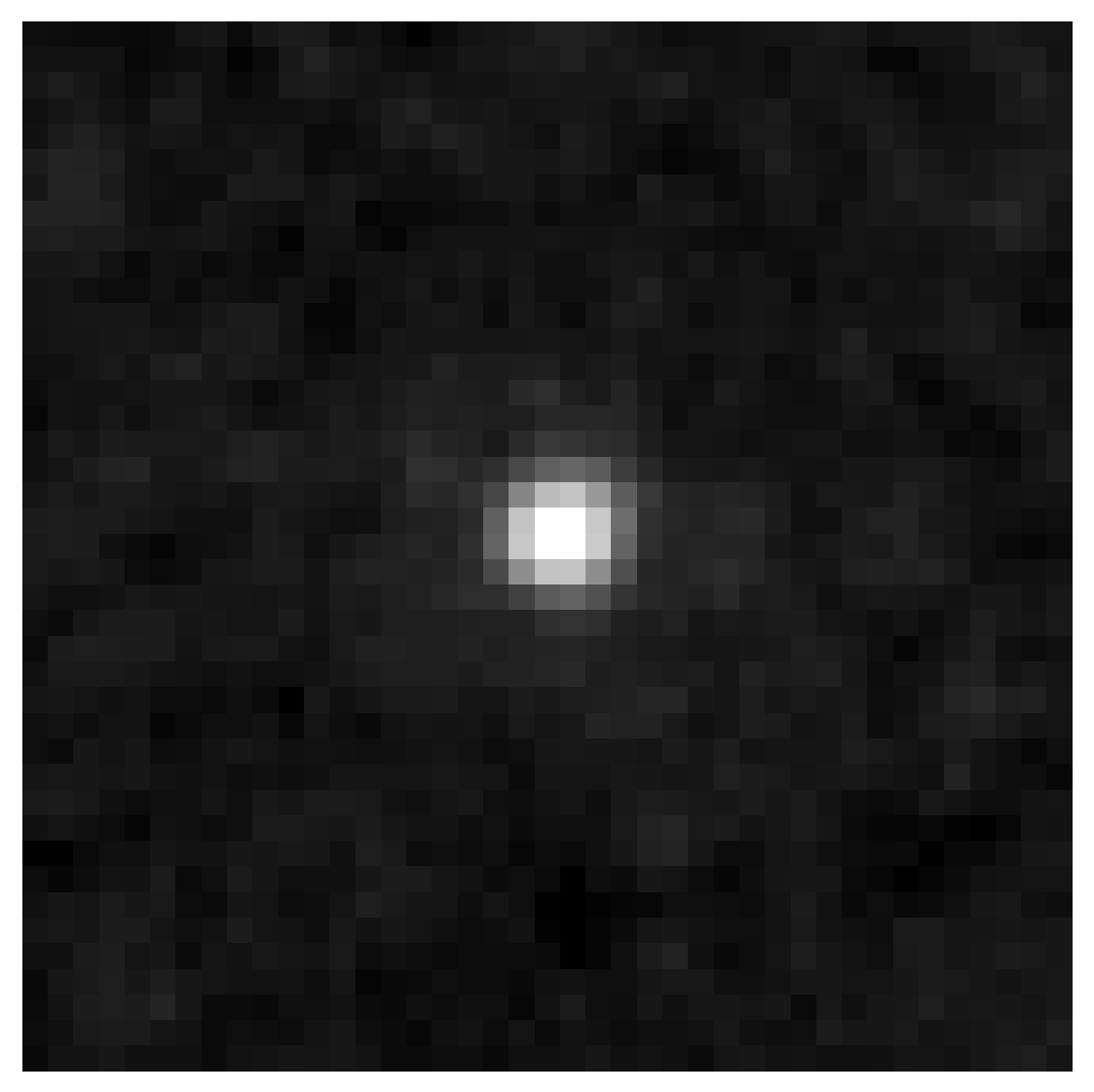}}
                \subfigure[359 -- 479~$\umu$Jy.]{
                \includegraphics[width=0.12\textwidth]{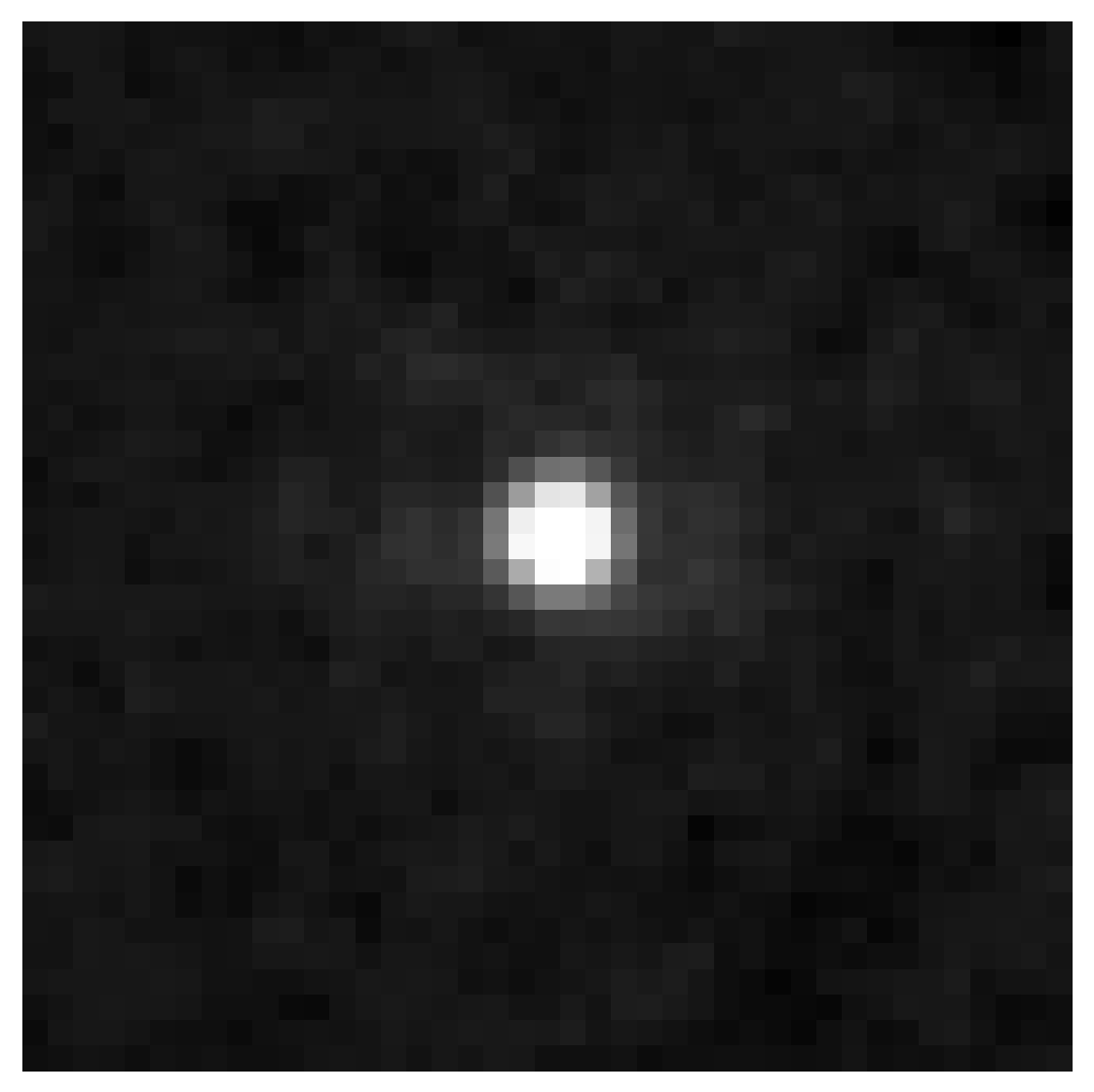}}
                \subfigure[479 -- 641~$\umu$Jy.]{ 
                \includegraphics[width=0.12\textwidth]{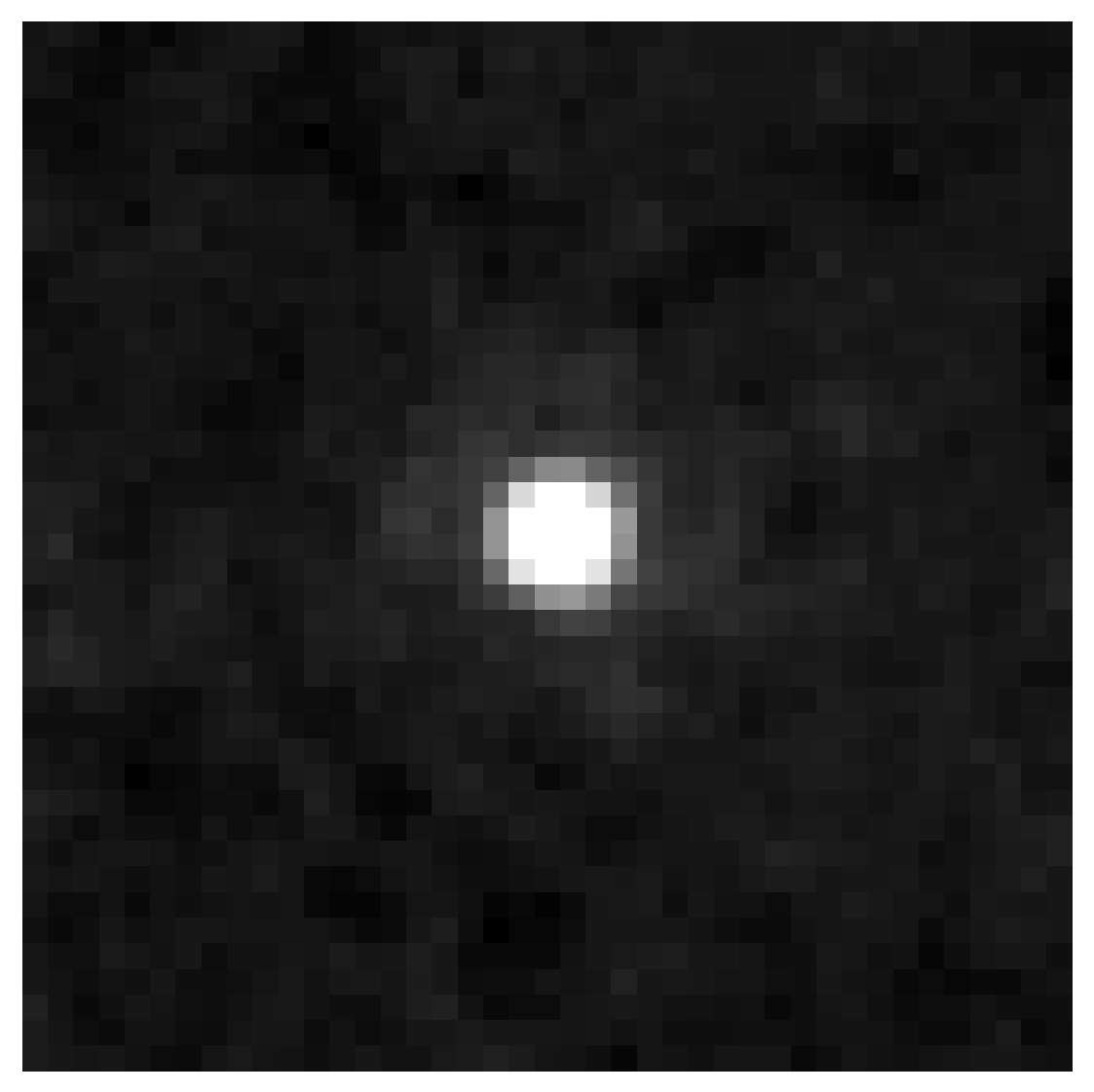}}
                \subfigure[641 -- 857~$\umu$Jy.]{
                \includegraphics[width=0.12\textwidth]{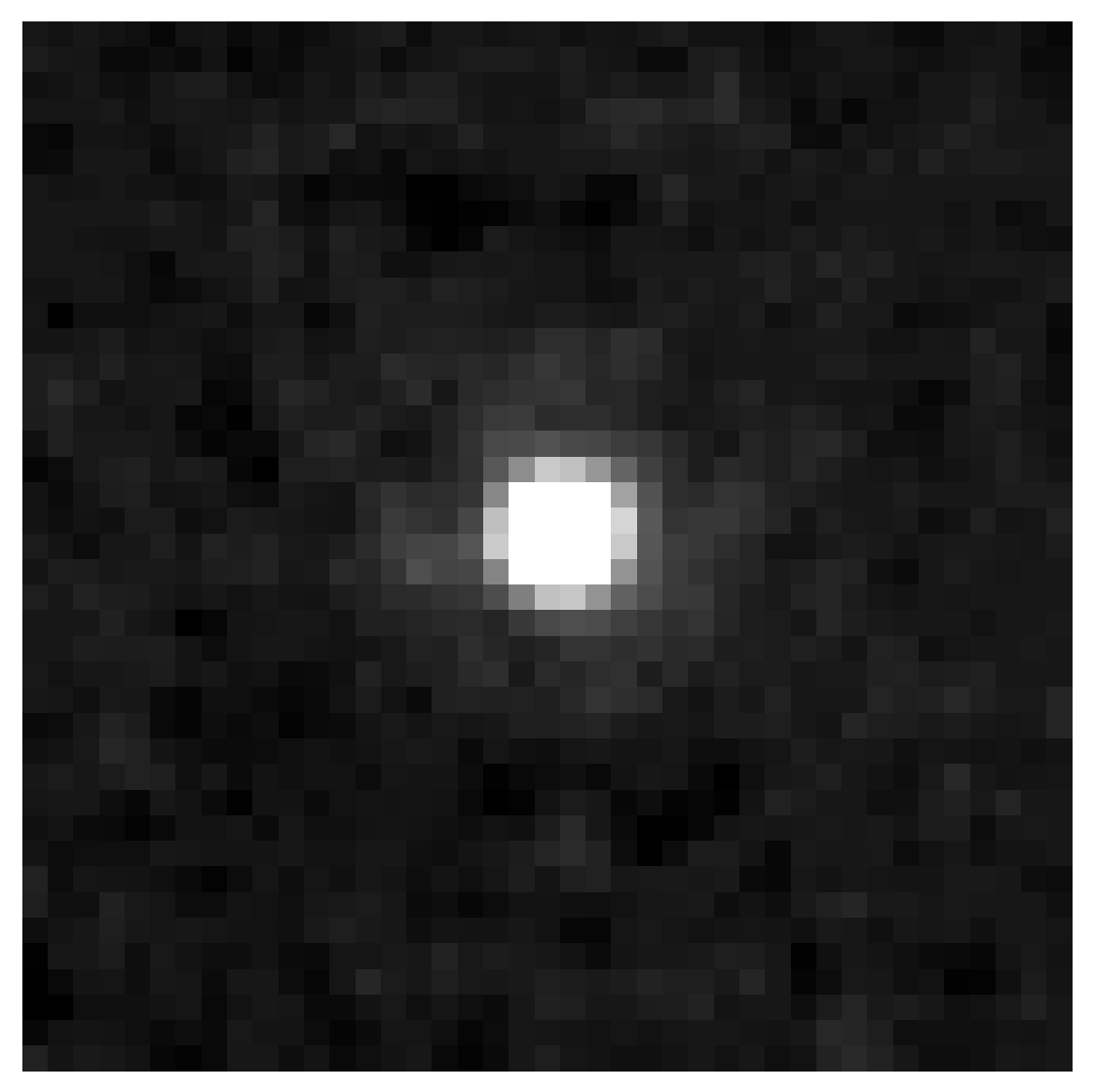}}}
  \caption{Median stacked 1.4-GHz radio images for the faintest six
  flux density bins.  All images have a size of
  $61.5\times61.5$~arcsec$^{2}$ ($41\times41$~pixel$^{2}$).  The
  24-$\umu$m flux density range of sources used to create the stacked
  image are given below each sub-image, with further details in
  Table~\ref{tab:imagedetails}.  The grey-scale ranges between $-2$
  and 20~$\umu$Jy~beam$^{-1}$.}
  \label{fig:stackedimagesmedian}
\end{figure*}

\begin{figure*}
  \centerline{\subfigure[150 -- 201~$\umu$Jy.]{
                \includegraphics[width=0.12\textwidth]{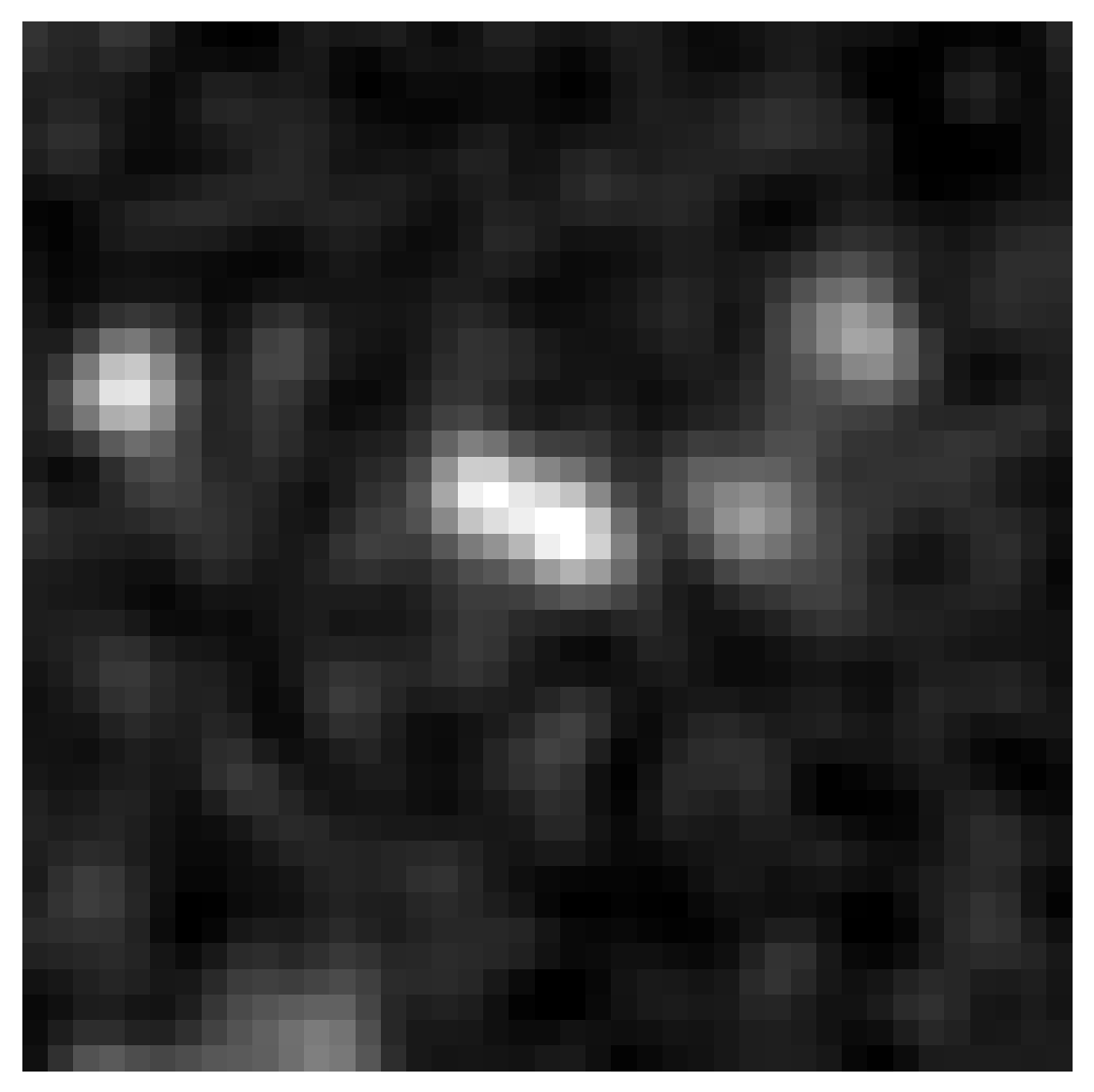}}
                \subfigure[201 -- 268~$\umu$Jy.]{
                \includegraphics[width=0.12\textwidth]{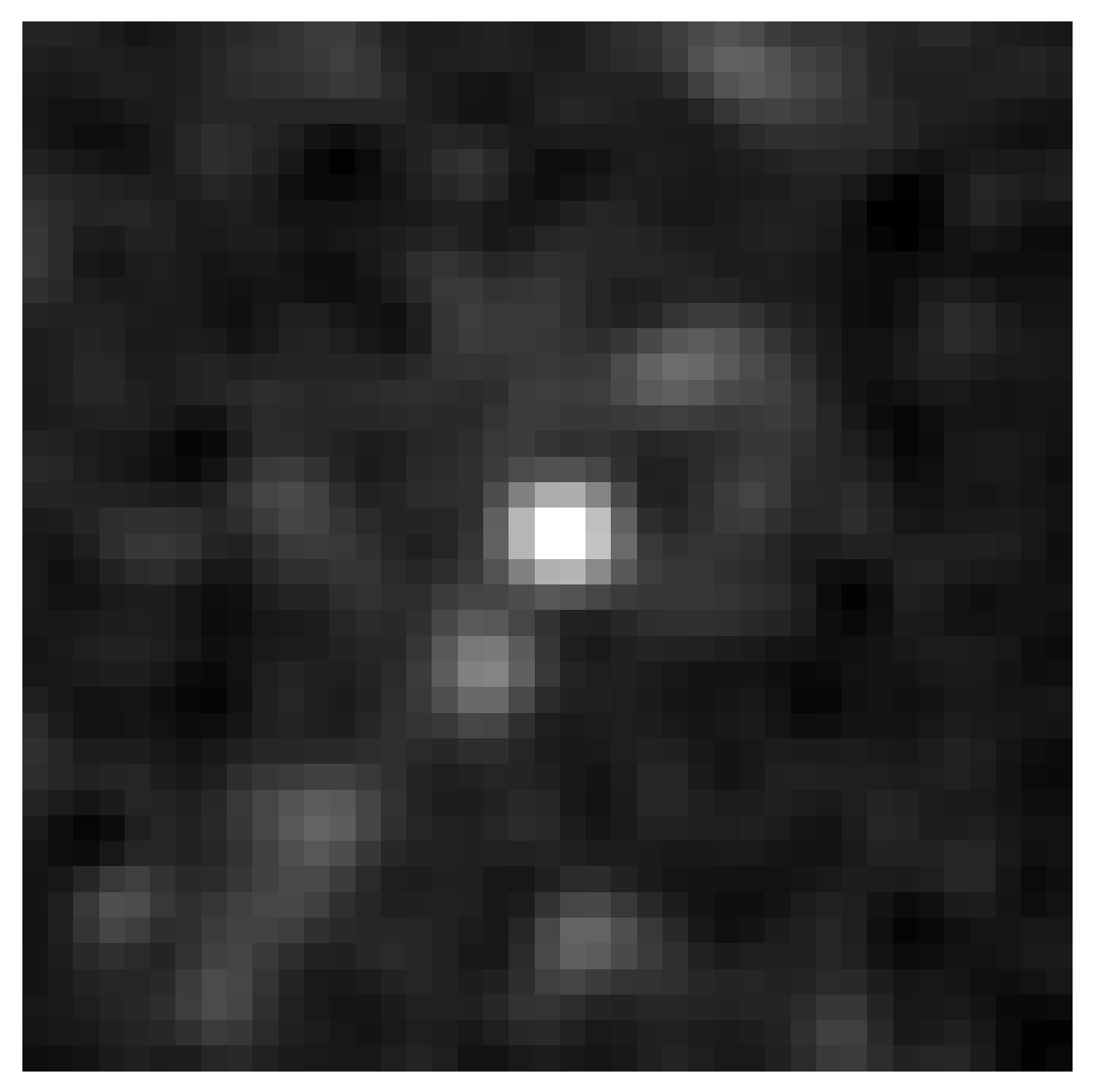}}
                \subfigure[268 -- 359~$\umu$Jy.]{
                \includegraphics[width=0.12\textwidth]{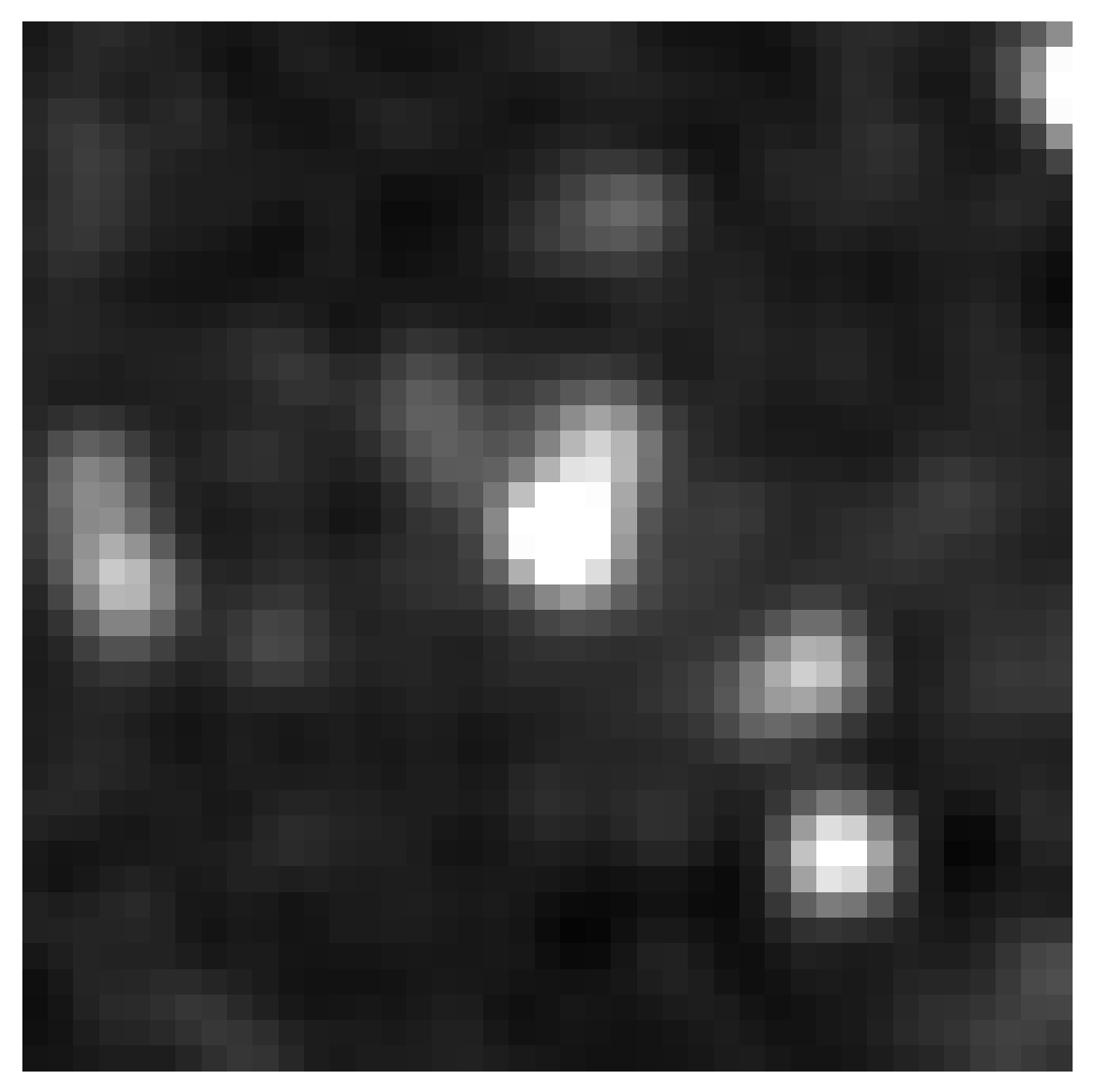}}
                \subfigure[359 -- 479~$\umu$Jy.]{
                \includegraphics[width=0.12\textwidth]{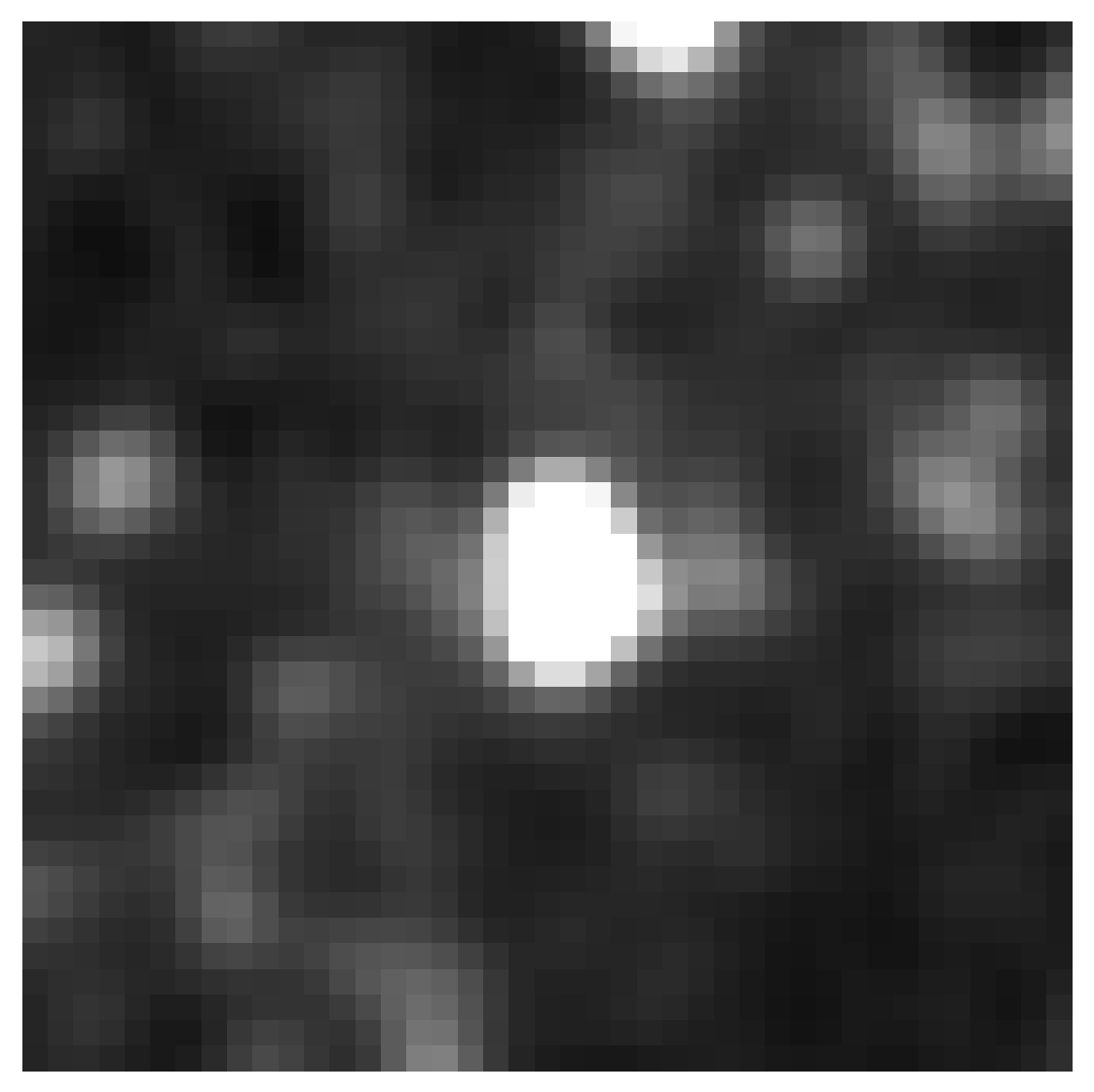}}
                \subfigure[479 -- 641~$\umu$Jy.]{ 
                \includegraphics[width=0.12\textwidth]{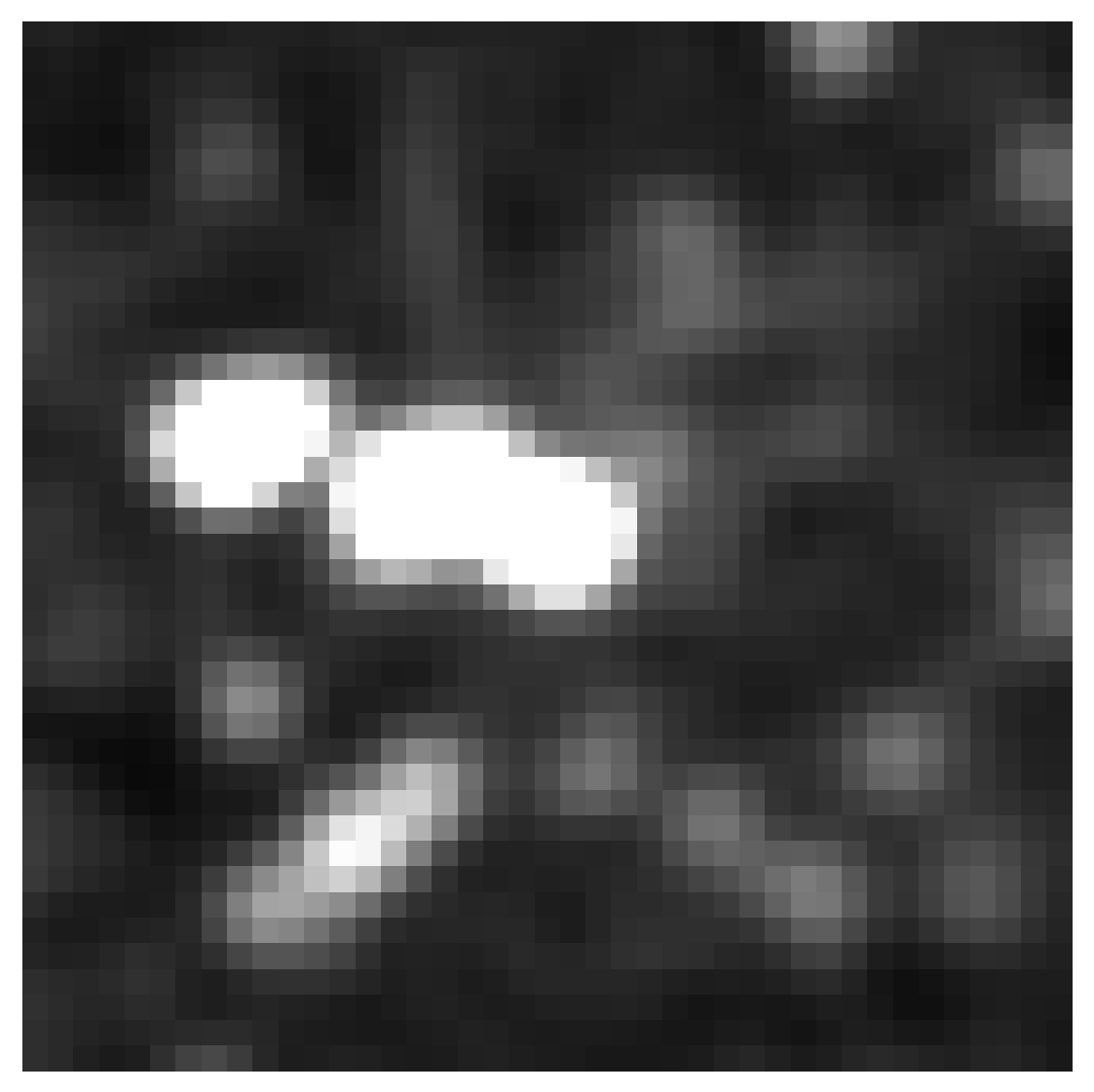}}
                \subfigure[641 -- 857~$\umu$Jy.]{
                \includegraphics[width=0.12\textwidth]{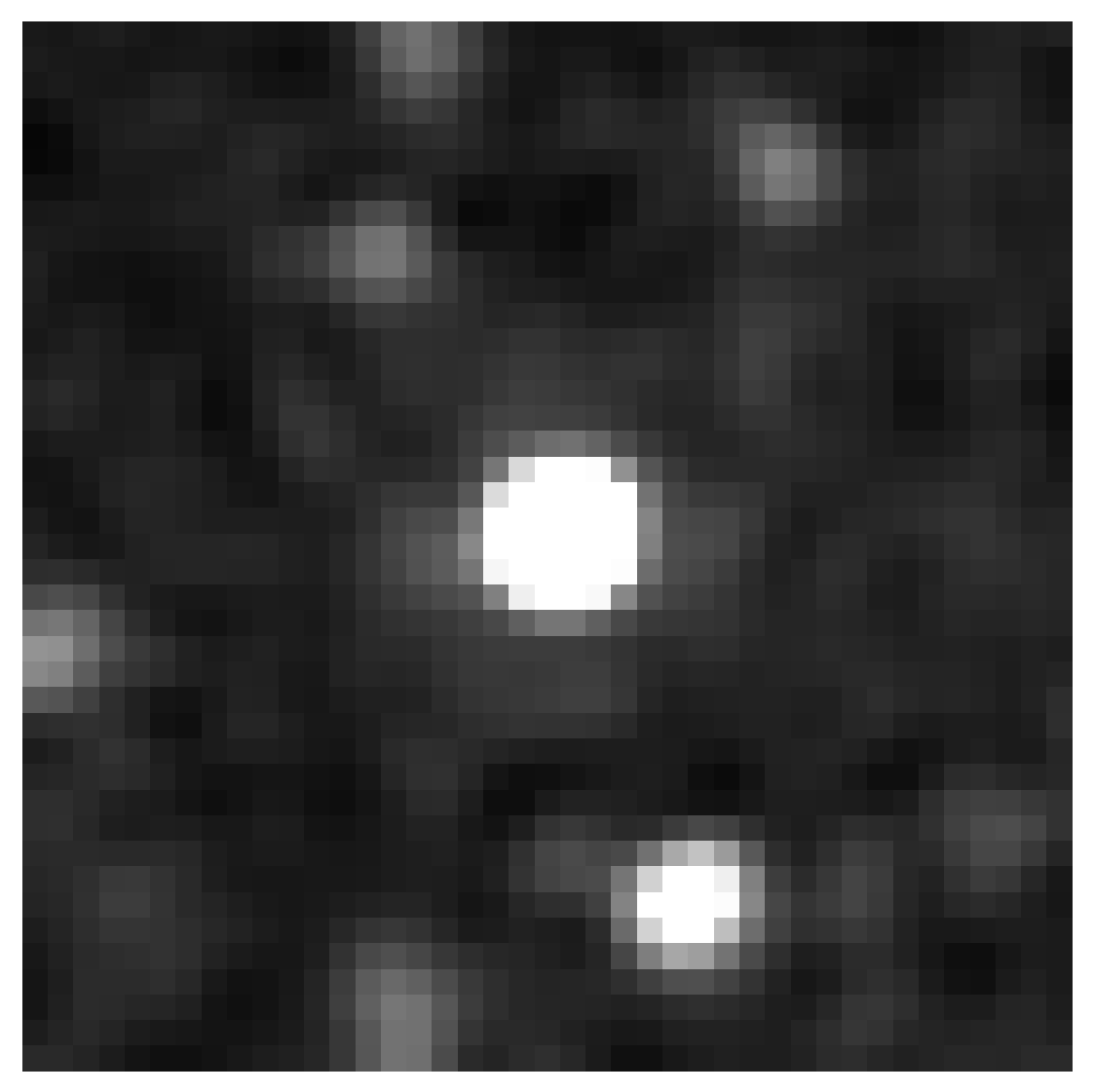}}}
  \caption{Mean stacked 1.4-GHz radio images for the same 24-$\umu$m
  flux density bins shown in Fig.~\ref{fig:stackedimagesmedian}.  As
  in Fig.~\ref{fig:stackedimagesmedian}, images are
  $61.5\times61.5$~arcsec$^{2}$ and the grey-scale ranges between $-2$
  and 20~$\umu$Jy~beam$^{-1}$, but the presence of a few bright sources
  in each stacked image make the resultant mean images no longer
  representative of the `typical' sources within a flux density bin.}
  \label{fig:stackedimagesmean}
\end{figure*}
The second method used in the literature to measure the flux density
of sources via stacking is to create a `stacked image' that is
representative of the average properties of sources within each
infrared flux density bin, and then calculate the radio flux density
directly from that image.  Small `cut-out' images are generated,
centred on each source within a flux density bin, and the value of
each pixel in the stacked image calculated from the mean or median of
the distribution of flux density that is seen from the equivalent
pixels in the $N$ cut-out images.

\begin{table}
  \begin{center}
  \caption{The infrared flux density range, number of sources in each
  bin $N$, median value of 1.4-GHz flux density $S_{1.4}$ and the
  noise level $\sigma$ for the six median stacked images shown in
  Fig.~\ref{fig:stackedimagesmedian}.}
  \label{tab:imagedetails}
    \begin{tabular}{ccccc}
\hline
Bin & $S_{24}$   &  $N$ & $S_{1.4}$ & $\sigma$\\
    & ($\umu$Jy)  &      & ($\umu$Jy) & ($\umu$Jy~beam$^{-1}$)\\
\hline
1   & 150 -- 201 &  466 &  21.2 & 1.21 \\
2   & 201 -- 268 &  855 &  25.2 & 0.80 \\
3   & 268 -- 359 & 2565 &  30.5 & 0.59 \\
4   & 359 -- 479 & 4583 &  39.0 & 0.42 \\
5   & 479 -- 641 & 2795 &  53.7 & 0.49 \\
6   & 641 -- 857 & 1476 &  72.2 & 0.64 \\
\hline
\end{tabular}
\end{center}
\end{table}

In order to compare this technique to the one described in
Section~\ref{sec:aperture}, we create $41\times41$~pixel$^{2}$
($61.5\times61.5$~arcsec$^{2}$) cut-out images which are centred on
each of the 24-$\umu$m sources within the xFLS field, and create mean
and median stacked images, using the same bins as in
Section~\ref{sec:aperture}.  The noise of the median stacked images
was calculated using {\sc Source Extractor} \citep{Bertin96}, and is a
good fit to a $1/\sqrt{N}$ relationship, although with an initial
noise level of 26~$\umu$Jy~beam$^{-1}$, 13~per~cent larger than the
value of 23~$\umu$Jy~beam$^{-1}$ quoted by \citet{Condon03}.  The bin
containing the most sources has a noise level of 424~nJy~beam$^{-1}$,
more than 50 times lower than the noise level of the VLA xFLS image,
and $\sim$ ten times lower than the most sensitive VLA observations to
date.

In Fig.~\ref{fig:stackedimagesmedian} we show the median stacked
images created for the six faintest bins, along with the range of
24-$\umu$m flux density that the bin covers.  The number of sources in
each bin, the total flux density of the stacked source and the noise
level of the stacked image are listed in Table~\ref{tab:imagedetails}.
There is a clearly-visible point source seen in each image, with a
circular appearance given by the $5\times5$~arcsec$^{2}$ resolution of
the original VLA mosaic.  In contrast to this,
Fig.~\ref{fig:stackedimagesmean} shows the mean stacked images for the
same six bins.  All images are noticeably noisier than their median
equivalents, and bright sources away from the centre of the cut-out
images have a much greater effect on the stacked images.  The noise
level of the mean images is $\sim1.5 - 3$ times the noise of the
median images, and the mean images are not representative of the
typical sources within each flux density bin, but are strongly biased
by a few bright radio sources.
 
We calculated the radio flux density of the stacked images, and
compared the results to the median radio flux density in each bin as
calculated in Section~\ref{sec:aperture}.  The median results are in
good agreement with each other, as is expected for point sources and a
Gaussian noise distribution -- the flux density measured from the
median image is equivalent to the median of the flux densities
measured from the individual images.  While the two methods for
estimating the median radio flux density give equivalent results,
creating stacked images means that the information on the IQR of the
flux density distribution is discarded.  We therefore use the median
binned flux density within an aperture for all future flux density
measurements in this work.

\subsection{Calculating median values of $q_{24}$}
\label{sec:VLAstackq}
\begin{figure}
  \centerline{\subfigure[$q_{24}$, calculated from all 14,820 sources
    in the 24-$\umu$m catalogue.  There is no significant variation in
    the median value of $q_{24}$ over the range of $S_{24}$ plotted.]{
    \includegraphics[width=0.45\textwidth]{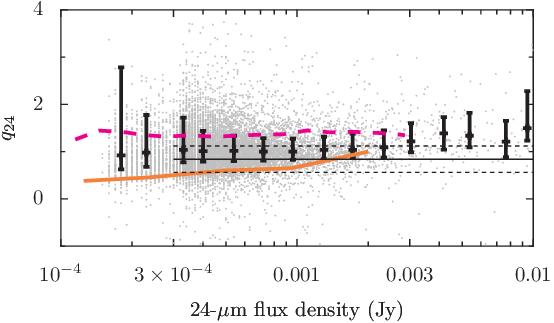}}}
    \centerline{\subfigure[$q_{24}$, calculated only from the 2,558
    sources that are also detected above $4\sigma$ at 1.4-GHz.  The
    median value of $q_{24}$ follows the trend of individual sources.
    The diagonal dashed line represents a constant radio flux density
    of $4\times23$~$\umu$Jy.  Bins containing fewer than 20 sources
    have been excluded from the figure.]{
    \includegraphics[width=0.45\textwidth]{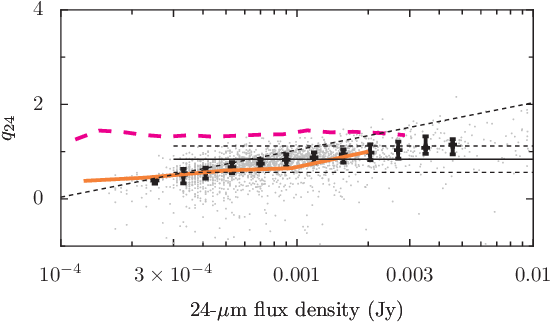}}}
  \caption{The effects of varying the selection criteria when
  calculating $q_{24}$.  Values of $q_{24}$ for individual sources are
  shown (grey dots), along with the median binned value of $q_{24}$
  (black points with errors denoting the IQR of the data).  Note that
  sources with an apparent radio flux density $< 0$ cannot be shown on
  these plots.  The decrease in the number of sources seen below a
  24-$\umu$m flux density of 300~$\umu$Jy is due to the smaller coverage
  area of the verification survey -- see Section~\ref{sec:techniques}
  for more details.  The values of $q_{24}$ found in the stacking
  experiments of \citet{Boyle07} (thick pink dashed line; CDFS field
  only, `all-source' data -- see Section~\ref{sec:stackingboyle} for
  further details) and \citet{Beswick08} (thick orange solid line) are
  plotted for comparison.  The horizontal black lines represent the
  mean value of $q_{24}$ of $0.84\pm0.28$ found by \citet{Appleton04}
  -- see text for more details.}
  \label{fig:VLAq24difference}
\end{figure}

The infrared / radio correlation can be quantified through the
logarithmic flux density ratio $q_{\rm IR}$ \citep{Appleton04}, where
\begin{equation}
  q_{\rm IR} = {\rm log}_{10} \left(\frac{S_{\rm IR}}{S_{1.4}}\right),
\end{equation}
and $S_{\rm IR}$ is the infrared flux density detected within either
the {\it Spitzer} 24-$\umu$m or 70-$\umu$m bands.  In
Fig.~\ref{fig:VLAq24difference}(a) the value of $q_{24}$ for each
source in the 24-$\umu$m catalogue is shown, along with the `median
binned' value of $q_{24}$, which is calculated directly from the
median values of infrared and radio flux density within each flux
density bin.  The error bars on $q_{24}$ come from the radio flux
density at the IQR for each flux density bin, and therefore represent
the same range as was shown in Fig.~\ref{fig:vlaflux}.  Since sources
with an apparent value of radio flux density that is negative (of
which there are many -- see Fig.~\ref{fig:vlaflux}) cannot be plotted
on this figure, the median value of $q_{24}$ does not follow the trend
of the individual plotted points directly -- this can be seen by the
asymmetric distribution range for the fainter flux density bins.

\citet{Appleton04} carried out the first study of the infrared / radio
correlation in the xFLS field, using a 1.4-GHz radio catalogue
\citep{Condon03}, and early 24-$\umu$m {\it Spitzer} data.  508 sources
were found with both 24-$\umu$m and 1.4-GHz detections.
\citet{Appleton04} calculated a mean value of $q_{24} = 0.84\pm0.28$
for these sources, which is shown on
Fig.~\ref{fig:VLAq24difference}(a) for comparison with our data.  Note
that the initial survey data which has been used for both the
\citet{Appleton04} study and the study described in this section is
the same.  It is clear that our stacking results give a value of
$q_{24}$ that is greater than the \citet{Appleton04} mean value
\citep[i.e.\ we find that sources are more radio-quiet than was found
by][]{Appleton04}, which is to be expected due to the fact that the
\citet{Appleton04} sample contained only those sources detected at both
frequencies, and is therefore subject to significant sample bias.

In order to test the effects of this bias, we created a catalogue of
1.4-GHz radio sources in the xFLS field, using the method described in
\citet{Garn07}, and requiring that sources could be detected at the
$4\sigma$ level.  We then repeated the stacking experiment described
above, using only the 2,558 24-$\umu$m sources which were also present
in this 1.4-GHz catalogue.  Fig.~\ref{fig:VLAq24difference}(b) shows
the individual values of $q_{24}$ for each source, calculated using
the method described in Section~\ref{sec:aperture}, along with the
median binned values of $q_{24}$.  The $4\sigma$ limit is shown on
this plot, although some sources have a measured flux density below
this limit (i.e.\ a value of $q_{24}$ above the dashed line) due to
the effects of noise on the aperture flux density measurements.  An
artificial variation of $q_{24}$ with infrared flux density is now
seen, although this is purely a selection effect and clearly
demonstrates the disadvantage of only studying the bright sources
which can be detected in radio images.  Now that the median values of
$q_{24}$ are being calculated from sources that can be clearly seen
above the noise, the binned data follows the trend of individual
sources and the flux density distribution is both narrower and more
symmetric.

A comparison of the \citet{Appleton04} results and the binned data
from detected sources in Fig.~\ref{fig:VLAq24difference}(b) shows good
agreement, with the binned results spanning the width of the
distribution found by \citet{Appleton04}.  This suggests that the
stacking method is not significantly biasing the radio flux density
which is measured for sources.  We overlay the values of $q_{24}$
found in the stacking experiments of \citet{Boyle07} and
\citet{Beswick08} on Fig.~\ref{fig:VLAq24difference} for comparison
with our data.  Note that the \citet{Beswick08} results are not
suffering from the selection effect which was described above,
resulting from only using detected sources, despite appearing to
follow a similar trend to our binned data in
Fig.~\ref{fig:VLAq24difference}(b) -- 80~per~cent of their infrared
sources can be detected above $3\sigma$ in their radio image, and all
sources have been used for the stacking experiment (see
Section~\ref{sec:stackingbeswick} for further details).  All three
stacking experiments find very different values for $q_{24}$, with the
sources in the study of \citet{Boyle07} being more radio-quiet than
our data, and the sources in the \citet{Beswick08} study being more
radio-loud.  A detailed comparison of our results with those of
\citet{Boyle07} and \citet{Beswick08} will be performed in later
sections.

\section{610-MHz stacking}
\label{sec:stack610}
\subsection{{\it Spitzer} extragalactic First Look Survey field}
The strength of the infrared / radio correlation can be quantified at
610~MHz in a similar way to at 1.4~GHz, using
\begin{equation}
  q'_{\rm IR} = {\rm log}_{10}\left(\frac{S_{\rm IR}}{S_{610}}\right),
\end{equation}
where $S_{610}$ is the radio flux density measured at 610~MHz.  There
are 13,812 24-$\umu$m sources located within the region covered by the
610-MHz GMRT image of the xFLS field \citep{Garn07}.  While the noise
level of the 1.4-GHz image is fairly uniform across the field, the
GMRT image shows more variation \citep[see Fig.~1 of][]{Garn07}.  In
order to test whether this varying noise affects the calculated values
of $q'_{24}$ we carried out the stacking procedure described in
Section~\ref{sec:aperture} on the GMRT image, using only those sources
within a radius of 25~arcmin from the centre of the mosaic.  This
analysis was repeated in 12.5~arcmin increments up to 1.25~deg --
increasing the radius had no significant effect on the median values
of $q'_{24}$, indicating that the varying noise level within this
image was not affecting the results.

\begin{figure}
  \centerline{\subfigure[Measured and median binned radio flux density
    for the 13,812 24-$\umu$m sources within the xFLS 610-MHz image.]{
    \includegraphics[width=0.45\textwidth]{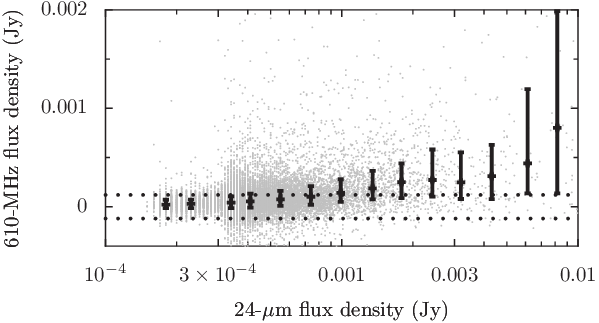}}}
    \centerline{\subfigure[Measured and median binned radio flux
    density for 13,812 random source positions, with each position
    assigned a 24-$\umu$m flux density from the catalogue at random.
    The radio flux density in each bin is consistent with zero.]{
    \includegraphics[width=0.45\textwidth]{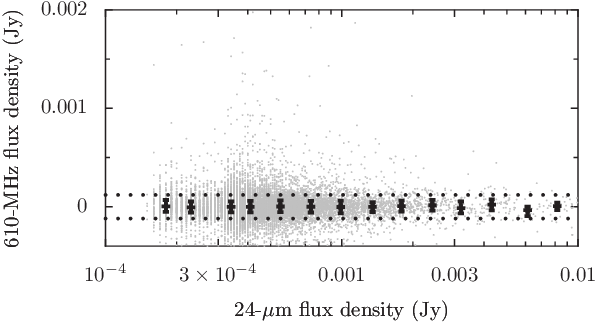}}}
  \caption{The 610-MHz radio flux density for each source in the
  24-$\umu$m catalogue, measured as in Section~\ref{sec:aperture} (grey
  dots), the values of $\pm4\sigma$ (horizontal black dotted lines)
  for the central region of the image, and the median values of flux
  density within each bin (black points with errors).  The error bars
  denote the size of the IQR of radio flux density in each bin.  The
  decrease in the number of sources seen below a 24-$\umu$m flux
  density of 300~$\umu$Jy is due to the smaller coverage area of the
  verification survey -- see Section~\ref{sec:techniques} for more
  details.}
  \label{fig:gmrtxflsflux}
\end{figure}

We varied the aperture size, and verified that a 4~arcsec aperture was
still appropriate for the GMRT image.  Fig.~\ref{fig:gmrtxflsflux}(a)
shows the 610-MHz flux density measured for each source, and the
median binned values, for comparison to Fig.~\ref{fig:vlaflux}(a).
Due to the increased and varying noise level across the 610-MHz image,
there is a greater dispersion in the individual values of flux density
being measured, and the IQR of radio flux density for each bin is
larger.  This is more clearly seen in Fig.~\ref{fig:gmrtxflsflux}(b),
where the flux density measured at 13,812 random source positions is
plotted -- since the noise level is not uniform, the individual flux
density measurements do not lie as tightly inside the $\pm4\sigma$
lines as before (where $\sigma=30$~$\umu$Jy~beam$^{-1}$, the noise at
the centre of the 610-MHz image).  The median of the distribution is
again consistent with having zero flux density in each bin.

\subsection{Stacking by source type in the SWIRE fields}
\label{sec:stacktype}
We have carried out no selection by source type so far in this work,
and have been stacking {\it all} 24-$\umu$m sources together, rather
than just those specifically associated with star formation.  The
24-$\umu$m xFLS catalogue of \citet{Fadda04} does not classify sources
by type, and it is possible that there may be a significant number of
stars or AGN in the sample, potentially affecting conclusions that are
being drawn for the star-forming galaxy population.  At 30~mJy the
contribution of stars and galaxies to the 24-$\umu$m source counts is
roughly the same, but galaxies rapidly dominate below this flux
density, and stars do not make up a significant fraction of the source
counts over the 24-$\umu$m flux density range that we are considering
\citep{Shupe08}.

We carried out source stacking in the three northern SWIRE fields,
using the GMRT 610-MHz images of the ELAIS-N1 \citep{Garn08EN1},
Lockman Hole \citep{Garn08LH} and ELAIS-N2 \citep{Garn09EN2} fields,
and obtaining positional information on infrared sources from the
photometric redshift catalogue of \citet{RowanRobinson08}.  While the
photometric catalogue is not completely unbiased, due to the
selection requirements and the need for sufficient photometry to
estimate redshifts, all sources within the catalogue have been
classified as either galaxies or AGN (with no stars being present).
The possibility that the xFLS sample is contaminated by significant
numbers of sources that are not star-forming galaxies can thus be
tested.

\begin{figure}
  \begin{center}
    \includegraphics[width=0.45\textwidth]{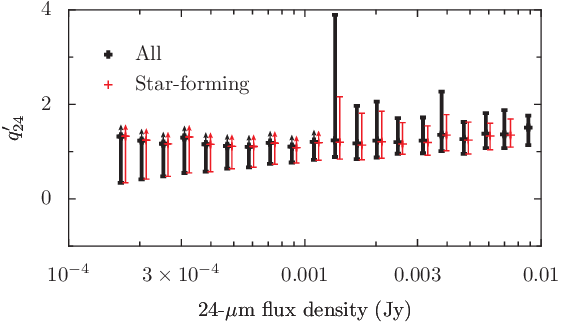}
    \caption{Median values of $q'_{24}$ for sources within the
    ELAIS-N1 field, along with the IQR of the data.  All sources from
    the \citet{RowanRobinson08} photometric catalogue are shown (thick
    black points), compared with sources that have been classified as
    star-forming (thin red points).  The star-forming sources have
    been displaced by a small amount along the x-axis for clarity, but
    would otherwise lie directly on top of the median value calculated
    from all sources.  No significant difference is seen between the
    two median values, although the distribution width is smaller for
    the star-forming sources, demonstrating that AGN contamination
    increases the uncertainty of the results.}
    \label{fig:en1q24types}
  \end{center}
\end{figure}

There are 48,882 entries in the photometric catalogue with 24-$\umu$m
flux densities between 150~$\umu$Jy and 10~mJy, within the region
covered by the ELAIS-N1 image.  Of these sources, 2,236 (4.6~per~cent)
have been classified as fitting AGN templates through their optical
and infrared photometry, although no information is available as to
which sources may have a significant amount of their radio emission
resulting from an AGN, but are not identified as such in the infrared.
Fig.~\ref{fig:en1q24types} shows a comparison between $q'_{24}$
calculated from all sources within the ELAIS-N1 field, and $q'_{24}$
from just those sources classified as star-forming. No significant
difference is seen in the median value, with the values for the
star-forming sources lying on top of the median values found when
using all sources in the catalogue.  The IQR decreases when only
star-forming sources are used, demonstrating that the AGN sources
increase the dispersion in the data, but that the median is resistant
to the presence of a small percentage of outliers.  The greater noise
level for the ELAIS-N1 survey compared with the xFLS 610-MHz survey
\citep[$\sim70$~$\umu$Jy~beam$^{-1}$ across the majority of the
image;][]{Garn08EN1} means that the IQR can permit negative radio flux
densities for some of the fainter infrared flux density bins -- where
this occurs, lower limits are shown for the IQR of $q'_{24}$ in
Fig.~\ref{fig:en1q24types}.

Repeating this test on the other two SWIRE fields, similar results
were found.  The presence of known AGN sources in the sample has very
little effect on the stacking results, and we conclude that
contamination from AGN sources in the xFLS sample will be unimportant.
However, all future stacking within the SWIRE fields will use only
those 24-$\umu$m sources identified as star-forming, with 46,646
sources in ELAIS-N1, 32,265 in ELAIS-N2 and 28,042 in the Lockman
Hole.

\section{Discussion}
\label{sec:stackingdiscussion}
\subsection{SWIRE and xFLS comparison}
\label{sec:errors}
\begin{figure}
  \begin{center}
    \includegraphics[width=0.45\textwidth]{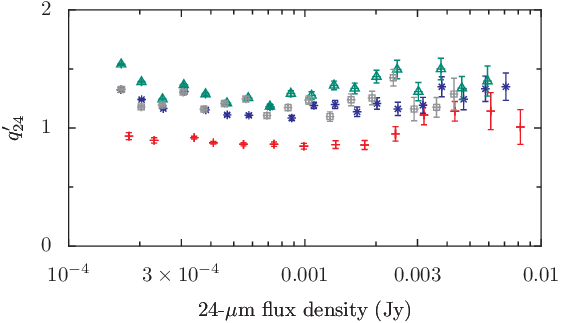}
    \caption{The median values of $q'_{24}$ from the xFLS (red upright
    crosses) and three SWIRE fields (ELAIS-N1 -- blue stars; ELAIS-N2
    -- green triangles; Lockman Hole -- grey squares).  The error bars
    represent the error on the value of the median radio flux density,
    rather than the IQR.}
    \label{fig:xflsswire610}
  \end{center}
\end{figure}

In Fig.~\ref{fig:xflsswire610} we show the 610-MHz stacking results
from the xFLS and SWIRE fields.  There is an apparent field-to-field
variation seen in the values of $q'_{24}$, with the xFLS data being
systematically lower by $\sim0.3$ than the SWIRE data.  The offset
affects all of the 24-$\umu$m flux density bins by approximately the
same amount.  The error estimates come from the uncertainty in median
radio flux density, which scales as $1/\sqrt{N}$ and is equivalent to
the error obtained from the noise level of a stacked image.  Due to
the dependence on $\sqrt{N}$, the fainter flux density bins (which
contain many more sources) appear to be much better constrained than
the brighter bins, where sources can actually be detected
individually.

There are a number of potential explanations for the offset which is
seen: 
\begin{enumerate}
\item A systematic offset in the flux density calibration in one or
  more of the radio images (Section~\ref{sec:radioflux}).
\item A systematic offset in the flux density calibration in one or
  more of the infrared catalogues (Section~\ref{sec:infraredflux}).
\item A bias being introduced into the stacking experiments
  due to sample incompleteness (Section~\ref{sec:stackingbias}).
\end{enumerate}

\subsection{Radio flux density calibration}
\label{sec:radioflux}
Any systematic flux density calibration offset in the radio surveys
would directly translate into a systematic offset in the calculated
value of $q_{24}$.  All of the GMRT surveys
\citep{Garn07,Garn08EN1,Garn08LH,Garn09EN2} were calibrated in the
same way, using observations of 3C48 or 3C286.  The relative flux
density calibration of the xFLS survey is accurate to 7~per~cent
\citep{Garn07}, and similar values apply to the three SWIRE survey
fields.  The effect of a constant calibration error is simple to
quantify -- if the measured radio flux density $S_{\rm m}$ equals
$f\times$ the true flux density $S_{\rm true}$, then the measured
value of $q_{\rm m}$ is related to the true value $q_{\rm true}$ via
\begin{equation}
  q_{\rm m} = q_{\rm true} - {\rm log}_{10}(f).
  \label{eq:fracq}
\end{equation}
A 7~per~cent flux density calibration error would lead to an error in
$q_{24}$ of $\pm0.03$ -- ten times too small to account for the
difference seen in Fig.~\ref{fig:xflsswire610}.

\begin{figure}
  \centerline{\includegraphics[width=0.45\textwidth]{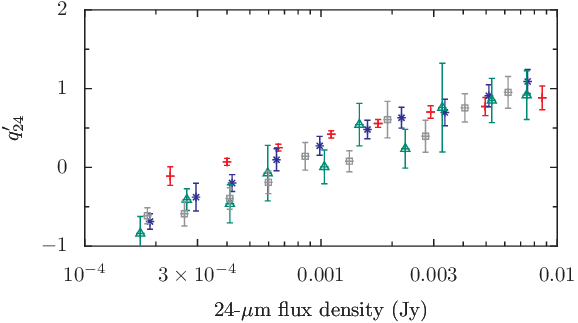}}
  \caption{A comparison of $q'_{24}$ for each of the GMRT fields,
  using the detected radio sources only, with error bars representing
  the error on the median value of radio flux density.  The colour
  scheme is the same as for Fig.~\ref{fig:xflsswire610}.  While the
  varying biases affect the values of $q'_{24}$ found for faint
  infrared flux densities, no systematic offset is seen between the
  values of $q'_{24}$ in the four fields for bright sources, in
  contrast to the result seen in Fig.~\ref{fig:xflsswire610} when all
  infrared sources are used.}
  \label{fig:gmrtalldetect}
\end{figure}

The existence of a systematic calibration offset between the radio
surveys can be tested for by considering the only those values of
$q'_{24}$ which are calculated from infrared sources with counterparts
in the GMRT catalogues from
\citet{Garn07,Garn08EN1,Garn08LH,Garn09EN2}.
Fig.~\ref{fig:gmrtalldetect} compares the median values of $q'_{24}$
from each of the GMRT images, using only those detected sources --
while the results for faint infrared flux density bins are affected
strongly by the same selection effects as were seen in
Fig.~\ref{fig:VLAq24difference}(b), no systematic offset is seen
between the median values of $q'_{24}$ for bright infrared sources, in
contrast to that seen in Fig.~\ref{fig:xflsswire610}.  The existence
of a systematic offset between the stacked values of $q'_{24}$ in the
xFLS and SWIRE fields, which is not seen in the values of $q'_{24}$
calculated from sources which are detected in the radio images, rules
out the possibility that the offset is a result of a difference in
flux density calibration between the radio surveys.  The median value
of $q'_{24}$ from bright, detected sources for all fields is $\sim1$
in the brightest 24-$\umu$m bin, which is in agreement with the value
found for the xFLS field using all 24-$\umu$m sources.

An error due to the radio flux density calibration can also be ruled
out by considering the radio source counts presented in
\citet{Garn08EN1} -- if the xFLS field was systematically brighter
than the SWIRE fields, then the 610-MHz differential source counts
would be expected to show a systematic offset between the xFLS and
SWIRE fields; no such effect was seen.  The GMRT source counts also
agree with GMRT counts made by other authors, confirming that the
relative flux density calibration of the surveys is accurate.  It
seems unlikely that any large-scale flux density calibration error can
be the cause of the discrepancy between the xFLS and SWIRE survey
data.

\subsection{Infrared flux density calibration}
\label{sec:infraredflux}
The 24-$\umu$m catalogues for the xFLS and SWIRE fields have been
created by separate authors (\citealp[xFLS;][]{Fadda04},
\citealp[SWIRE;][]{RowanRobinson08}).  While it is equally unlikely
that any large-scale flux density calibration error in the infrared
could be responsible for the systematic effect which is seen, this can
be tested through a stacking experiment in the xFLS and SWIRE fields
using 70-$\umu$m sources.

\begin{figure}
  \begin{center}
    \includegraphics[width=0.45\textwidth]{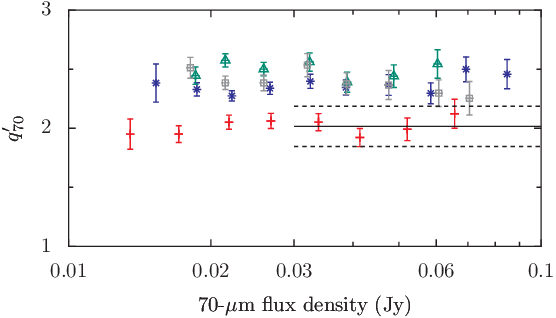}
    \caption{Median values of $q'_{70}$, along with errors in the
    median, taken from the four GMRT surveys.  The colour scheme is
    the same as for Fig.~\ref{fig:xflsswire610}.  The horizontal
    black lines represent the mean value of $q_{70} = 2.16\pm0.17$
    found by \citet{Appleton04}, converted to a 610-MHz value using
    Equation~\ref{eq:q1.4610} and $\alpha=0.4$ -- see
    Section~\ref{sec:stackingbias}.}
    \label{fig:q70stack}
  \end{center}
\end{figure}

There are 601 sources in the \citet{Frayer06} xFLS catalogue contained
within the 610-MHz image, and 1028, 597 and 687 sources from the
\citet{Surace05} SWIRE catalogues within the ELAIS-N1, ELAIS-N2 and
Lockman Hole images.  We carried out the same stacking procedure,
using 70-$\umu$m source positions, in order to compare the results to
the 24-$\umu$m data.  Fig.~\ref{fig:q70stack} shows the $q'_{70}$
values for the four fields.  The error bars on $q'_{70}$ are greater
than before, due to the decreased number of sources in the infrared
catalogues.  The same field-to-field variation is seen, with a similar
systematic offset of $\sim0.3$ between the xFLS and SWIRE median
values of $q'_{70}$, and no variation in the value of $q'_{70}$ with
$S_{70}$ is seen over the range 10 -- 100~mJy for any of the survey
fields.  The similar offset seen in $q'_{24}$ and $q'_{70}$ implies
that the offset is not a result of a difference between the
\citet{Fadda04} and \citet{RowanRobinson08} 24-$\umu$m catalogues.  

We overlay the \citet{Appleton04} result of $q_{70} = 2.16\pm0.17$ on
Fig.~\ref{fig:q70stack}, converted to a 610-MHz value using
Equation~\ref{eq:q1.4610} and $\alpha=0.4$ (see
Section~\ref{sec:stackingbias}).  The lower sensitivity of the
70-$\umu$m data compared with the 24-$\umu$m observations ($\sim30$~mJy
compared with $\sim0.3$~mJy) makes the sample bias that occurs from
only using sources that can be detected in the radio less significant
than at 24-$\umu$m.  As was seen at 24-$\umu$m, the \citet{Appleton04}
result agrees well with the xFLS stacking experiment, although our
data covers a greater range of 70-$\umu$m flux density than
\citet{Appleton04}, as the \citet{Frayer06} source catalogue of the
field was not yet available.  The good agreement between our stacked
$q'_{70}$ values and the \citet{Appleton04} value is further evidence
that the stacking procedure is not significantly biasing the radio
flux density measurements.

\subsection{Stacking bias}
\label{sec:stackingbias}
\citet{White07} carried out a stacking experiment using the FIRST
survey, and ran simulations to test how well their stacking experiment
could recover the flux density of sources below the noise level.  They
placed four artificial sources in their {\it uv} data for each of 400
FIRST observations, and re-reduced the images to fully simulate all of
the potential sources of bias that may enter into the data reduction
procedure.  They found that some of the initial flux density could not
be recovered from stacked images, and attribute the loss to an effect
equivalent to `CLEAN bias' \citep[e.g.][]{Becker95}, despite the fact
that sources below the survey threshold {\it `by definition, have not
been CLEANed'} \citep[although see Section~\ref{sec:stackingboyle} for
a discussion of the stacking simulation carried out by][]{Boyle07}.
However, the second technique that \citet{White07} used to test for a
loss of flux was to take existing deeper observations of an area
covered by FIRST (the VLA xFLS survey) to provide positions and flux
densities for sources below the FIRST detection threshold, and then
stack sources using the prior knowledge of their radio flux density.
They found that their stacked images were able to recreate 71~per~cent
of the true flux density of sources, for all flux density bins.  If a
constant fraction $f$ of flux is being recovered for all sources, then
the required correction factor is also just a constant fraction (in
this case 1/0.71 = 1.41) for each infrared flux density bin, rather
than varying in a complex manner from source to source.  The overall
effect of this bias would be to alter the absolute value of $q_{24}$
which is measured, but without changing its dependence on infrared
flux density.  From Equation~\ref{eq:fracq}, the correction factor
found by \citet{White07} would have led to a calculated value which is
0.15 above the true value of $q_{24}$.  In order to obtain an offset
of 0.3, the correction factor would have to be ${\rm log_{10}}(0.3)
\simeq 2$, with only half of the radio flux density of stacked sources
being correctly measured.

The linearity of any potential stacking bias can be directly testing
within the xFLS field, where the same 24-$\umu$m catalogue has been
used to stack both 1.4-GHz and 610-MHz sources.  Any non-linear form
of bias would be expected to affect the two radio images in different
ways, due to the separate telescope arrays used to take the
observations, the independent data reduction methods, and the
different amount of CLEANing.  The relationship between $q_{24}$ and
$q'_{24}$, taking a single spectral index $\alpha$ to represent the
whole population\footnote{We define $\alpha$ such that the variation
of flux density $S_{\nu}$ with frequency $\nu$ is
$S_{\nu}=S_{0}\nu^{-\alpha}$.}, is
\begin{equation}
  q_{24} = q'_{24} - {\rm
  log}_{10}\left(\frac{S_{1.4}}{S_{610}}\right) = q'_{24} +
  0.36\alpha.
\label{eq:q1.4610}
\end{equation}

\begin{figure}
  \begin{center}
    \includegraphics[width=0.45\textwidth]{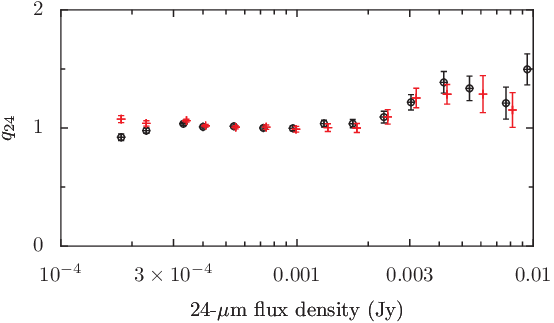}
    \caption{A comparison between $q_{24}$ obtained from the 1.4-GHz
    image of the xFLS field (black circles) and $q'_{24}$ from the
    610-MHz image of the same region, shifted to a 1.4-GHz value
    through the use of Equation~\ref{eq:q1.4610} and $\alpha=0.4$ (red
    upright crosses).  Error bars represent the error on the median
    value of radio flux density.  The shape of the two profiles is
    consistent across the full range of infrared flux density, ruling
    out any significant instrument-dependent or data
    reduction-dependent biases.}
    \label{fig:xflsvlagmrt}
  \end{center}
\end{figure}

By comparing the binned values of $q_{24}$ and $q'_{24}$ found in the
xFLS field, we are able to obtain a single spectral index,
$\alpha=0.4$, which represents the linear shift required to convert
between the stacked 610-MHz and 1.4-GHz flux densities.
Fig.~\ref{fig:xflsvlagmrt} shows the values of $q_{24}$ calculated
from the xFLS 1.4-GHz image, and the values of $q'_{24}$ calculated
from the xFLS 610-MHz image, converted to 1.4-GHz values using
$\alpha=0.4$.  The shape of the $q_{24}$ and (shifted) $q'_{24}$ data
are consistent across the full range of infrared flux density, with
the exception of the faintest flux density bin.  While the value of
$\alpha$ has been chosen in order to overlay the $q_{24}$ and
$q'_{24}$ profiles, making the absolute values shown in
Fig.~\ref{fig:xflsvlagmrt} not independent of each other, the
consistent shape of the $q_{24}$ and $q'_{24}$ profiles in
Fig.~\ref{fig:xflsvlagmrt} demonstrates that any stacking bias which
is present must be a simple fraction of the true flux density, as was
found by \citet{White07}, and rules out any significant non-linear
flux-dependent stacking bias.

\subsection{Comparison with the results of \citet{Boyle07}}
\label{sec:stackingboyle}

The stacking experiment of \citet{Boyle07} is similar to that
presented in this work, also using 24-$\umu$m sources detected in the
SWIRE survey, in the {\it Chandra} Deep Field South (CDFS) and
ELAIS-S1 fields.  The infrared observations of the ELAIS-S1 field were
much less sensitive than the CDFS observations, and the results from
the two fields were broadly similar, so we only compare our results to
those from the CDFS field.

The \citet{Boyle07} radio data came from a 1.4-GHz Australia Telescope
Compact Array (ATCA) survey, with an rms noise level of
30~$\umu$Jy~beam$^{-1}$ and resolution of $11\times5$~arcsec$^{2}$
\citep{Norris06}.  The {\it Spitzer} catalogue of the CDFS field
contains $\sim$12,000 sources above an infrared $5\sigma$ flux density
limit of 100~$\umu$Jy (while the ELAIS-S1 survey only contains
$\sim$2,000 sources above a flux density limit of 400~$\umu$Jy).
Median stacked radio images were created from the binned CDFS sources,
and the peak flux density of these images was taken as their estimate
of the radio flux density.  They select data in two ways -- an
`all-source' sample, where all the data was used, and a `quiet-source'
sample, where cut-out images that had a central pixel with radio flux
density of $>100$~$\umu$Jy were not used to create the stacked image.
Due to the exclusion of visibly-present radio sources, the
`quiet-source' stack suffers from a systematic bias in the brighter
flux density bins, and underestimates the radio flux density for the
stacked sources (as can be seen in their Fig.~4) -- for this reason,
all comparisons to \citet{Boyle07} data in this work are with the
`all-source' sample.

\citet{Boyle07} found a value of $q_{24}=1.39$, and no significant
variation with $S_{24}$ over the range 100 -- 2800~$\umu$Jy.
Fig.~\ref{fig:q24all} shows the values of $q_{24}$ found by
\citet{Boyle07} in the CDFS, and the values of $q'_{24}$ from the
three 610-MHz SWIRE fields, converted to 1.4-GHz values using the
spectral index of $\alpha=0.4$ found from the xFLS comparison in
Section~\ref{sec:stackingbias}.  The \citet{Boyle07} results agree
with the general trend of the 610-MHz SWIRE results, although the data
in individual flux density bins are not always completely consistent.
The errors on the fainter flux density bins from the 610-MHz SWIRE
fields are significantly smaller than the \citet{Boyle07} data, due to
the increased number of 24-$\umu$m sources present in the
\citet{RowanRobinson08} data compared with the earlier
\citet{Surace05} SWIRE catalogues used by \citet{Boyle07}.

\citet{Beswick08} argued that some of the discrepancy between their
results ($q_{24} = 0.48$; see Section~\ref{sec:stackingbeswick}) and
those of \citet{Boyle07} could be due to a systematic under-estimation
of the radio flux density by up to a factor of 2 by the ATCA for
sources at low flux densities.  This would lead to an over-estimation
of $q_{24}$ by \citet{Boyle07} of $\sim0.3$, reducing their value to
$\sim1.1$ in the faint 24-$\umu$m flux density bins (and therefore
agreeing with our xFLS data).  However, our findings are not
consistent with this interpretation, with our SWIRE stacking results
agreeing with the \citet{Boyle07} data after using a conversion
between 610-MHz and 1.4-GHz results calculated from the xFLS field,
independent of the SWIRE data.  \citet{Norris06} found through
comparing their radio data to a more sensitive survey of the
GOODS-South portion of the region \citep{Afonso06} that the ATCA flux
densities were only being under-estimated by about 14~per~cent, which
would lead to an over-estimation of the \citet{Boyle07} value of
$q_{24}$ by $\sim0.07$ -- this amount of variation is within the
typical field-to-field dispersion seen from the three 610-MHz SWIRE
surveys, and cannot be ruled out.  While any underestimation of ATCA
radio flux densities has been shown to be small at high flux density,
this effect may be more significant for lower flux density sources --
see e.g.\ \citet{Prandoni00} -- which would affect the faint stacked
sources ($<130$~$\umu$Jy) studied by \citet{Boyle07} more significantly
than the brighter discrete objects ($>200$~$\umu$Jy) discussed by
\citet{Norris06}.  It remains unclear as to the potential size or
cause of any systematic underestimation of radio flux densities in
ATCA observations.

\citet{Boyle07} recognized that their value of $q_{24}=1.39$ was
unexpectedly high -- {\it `greater than any modelled SED'}, and so
carried out simulations to test for any effects such as the stacking
bias described in Section~\ref{sec:stackingbias}.  They inserted
sources into their image, and confirmed that they were able to
recreate the median value of the source population through the use of
stacking.  More importantly in the context of this work, they inserted
sources into their {\it uv} data and reprocessed the images to fully
simulate all of the effects which could be biasing the stacked flux
density values.  Their stacking measurements gave them `{\it an
essentially identical result}' to simulations in the image plane, and
to their original data, which suggests that any stacking bias in their
experiment must be small.

The fact that the same conversion factor which was found for the xFLS
data (of $\alpha=0.4$) also allows the \citet{Boyle07} and 610-MHz
SWIRE data to agree implies that there can be little stacking bias in
the 610-MHz surveys.  Any significant bias which only applied to the
GMRT xFLS image would lead to a value of $\alpha$ which would not be
suitable for the SWIRE fields, contrary to what is seen.  We do not
believe that a stacking bias can be the cause of the observed
discrepancy between the xFLS and SWIRE results.

\subsection{Comparison with the results of \citet{Beswick08}}
\label{sec:stackingbeswick}
\begin{figure}
  \begin{center}
    \includegraphics[width=0.45\textwidth]{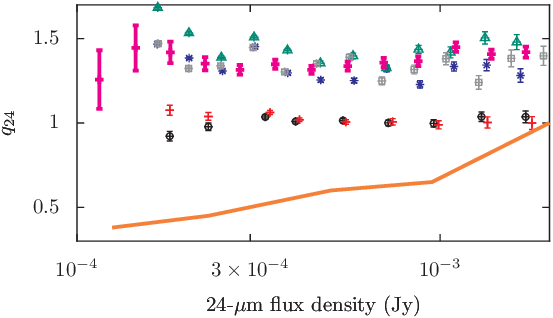}
    \caption{A comparison of all of the stacking results discussed in
    this work, using data shifted to 1.4~GHz where appropriate through
    the use of Equation~\ref{eq:q1.4610} and $\alpha=0.8$.  The
    stacking results from this work are given by thin data points with
    error bars, with the VLA xFLS image (black circles), GMRT xFLS
    image (red upright crosses), GMRT ELAIS-N1 image (blue stars),
    GMRT ELAIS-N2 image (green triangles), and GMRT Lockman Hole image
    (grey squares) being shown, along with results from the
    \citet{Boyle07} ATCA CDFS image (thick pink points; `all-source'
    data) and the \citet{Beswick08} MERLIN+VLA HDF-N image (thick
    orange solid line).  All error bars represent the error on the
    median value of radio flux density.}
    \label{fig:q24all}
  \end{center}
\end{figure}

\citet{Beswick08} performed their stacking experiment on a sample of
377 24-$\umu$m sources detected with {\it Spitzer} in the {\it Hubble}
Deep Field-North (HDF-N), 303 of which can be seen above $3\sigma$ in
a deep Multi-Element Radio-Linked Interferometer Network (MERLIN) and
VLA image of the field with rms noise level of 3.6~$\umu$Jy~beam$^{-1}$
and resolution of 0.4~arcsec.  They measured radio flux densities
using both of the techniques described earlier, and obtained
comparable results with the two methods.  \citet{Beswick08} found a
median value of $q_{24} = 0.48$ for all sources, and a `{\it tentative
deviation}' in the infrared / radio correlation at very low 24-$\umu$m
flux densities, with the fainter sources having lower values of
$q_{24}$.  Fig.~\ref{fig:q24all} shows all of the stacking results
presented in this work, converted to a 1.4-GHz value where
appropriate, along with those from \citet{Boyle07} and
\citet{Beswick08}.  In order to display the data more clearly we focus
on the 24-$\umu$m flux density range between 0.1 and 2~mJy.

There is a clear discrepancy between the results of \citet{Beswick08},
the xFLS results, and the results from the SWIRE surveys.  The most
noticeable effect is that the radio surveys with the lowest noise
levels show the lowest values of $q_{24}$ -- the \citet{Beswick08}
survey is the deepest, at 3.6~$\umu$Jy~beam$^{-1}$, while the xFLS
surveys are the next most sensitive, and the SWIRE fields are the
least sensitive.  However, the noise levels of the VLA xFLS survey
(23~$\umu$Jy~beam$^{-1}$) and the ATCA CDFS survey
(30~$\umu$Jy~beam$^{-1}$) are much more similar than the variation in
noise between the 610-MHz xFLS and ELAIS-N2 images
(30~$\umu$Jy~beam$^{-1}$ and 80~$\umu$Jy~beam$^{-1}$), and yet the
stacking results from those fields are still clearly inconsistent with
each other, suggesting that this cannot be the only factor which is
important.

\begin{figure}
  \centerline{\subfigure[xFLS -- 13812 sources.]{
    \includegraphics[width=0.2\textwidth]{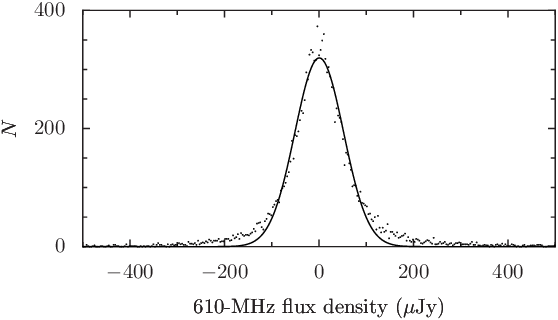}}
              \subfigure[ELAIS-N1 -- 46646 sources.]{
    \includegraphics[width=0.2\textwidth]{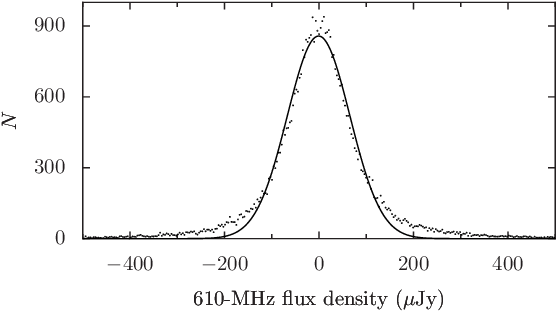}}}
  \centerline{
  \subfigure[ELAIS-N2 -- 32265 sources.]{
    \includegraphics[width=0.2\textwidth]{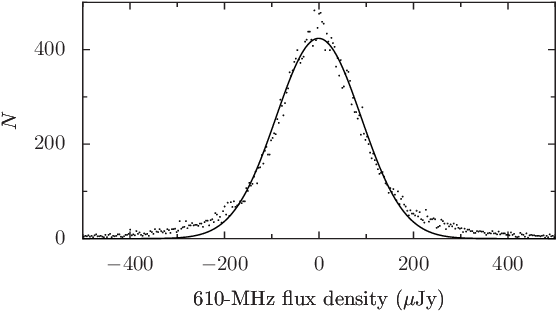}}
              \subfigure[Lockman Hole -- 28042 sources.]{
    \includegraphics[width=0.2\textwidth]{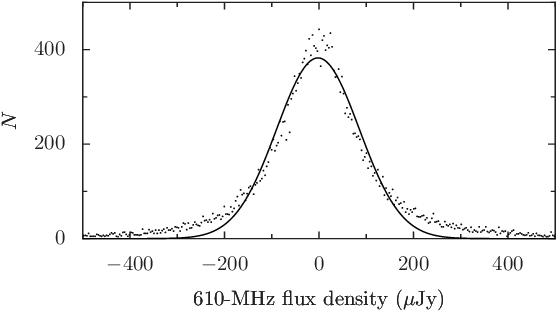}}}
  \caption{Histograms of the flux density recorded within $N$ random
  apertures in each of the 610-MHz surveys.  No aperture correction
  factor has been applied to the measured flux densities.}
  \label{fig:randomhistograms}
\end{figure}

In order to test whether the stacking results are being biased by the
noise level of the radio image, the distribution of flux density
within $N$ random apertures was calculated for each of the GMRT
images, and shown in Fig.~\ref{fig:randomhistograms}.  No aperture
correction was applied to these flux density measurements, since
applying the aperture correction only broadens each distribution by a
factor of between 1.4 and 1.5, and does not affect the conclusions of
this section.  The slightly varying noise levels near to the edge of
each image mean that a single Gaussian can not completely represent
the flux density distribution, but there is no evidence for any
asymmetry or significant non-Gaussianity in the measured values of
flux density for any of the fields.  The effects of noise on the
stacking procedure will not systematically bias the measured values of
radio flux density, and can not be responsible for the systematic
difference in $q'_{24}$ which is seen between fields.

There may be a significant number of radio-bright AGN in the HDF-N
sample, which would lead to an decrease in $q_{24}$.  Any AGN
contamination would have to be large, since the 4.6~per~cent of AGN
sources within the ELAIS-N1 field made essentially no difference to
the calculated value of $q'_{24}$ in Section~\ref{sec:stacktype}, and
the median is naturally resistant to being affected by small numbers
of outlier sources.  All of the stacking studies presented here are
infrared-selected, and the population would therefore be expected to
be dominated by star-forming systems -- this makes it unlikely,
although not impossible, that large amounts of AGN contamination will
be a problem.  \citet{Biggs06} have shown that the differential number
counts of radio sources within the HDF-N are consistent with source
counts from other deep observations -- although this does not provide
any information directly about the brightness of infrared-selected
sources, it does show that the HDF-N is not an anomalously
radio-bright region of sky.

80~per~cent of the \citet{Beswick08} sample are detected in the radio
above $3\sigma$, making it much more complete than the samples
described in this work (the deepest of which has 2,558/14,820 =
17~per~cent radio detections above $4\sigma$).  It is important to
note that we have applied {\it no} radio detection criteria to the
stacked samples (a major advantage of stacking experiments), so no
bias should be entering the analysis through incompleteness --
however, if the `stacking bias' identified by \citet{White07} did vary
with flux density, or S/N, then a more incomplete sample may lead to a
greater value of $q_{24}$ being found.  Either way, in order to change
a value of $q_{24}=0.48$ to 1.39 (from Beswick to Boyle) would require
from Equation~\ref{eq:fracq} that only 12~per~cent of the true radio
flux density was being measured by \citet{Boyle07}, while
\citet{Beswick08} were measuring the full value -- it seems highly
unlikely that such a large error could be occurring.  There may be
some stacking bias taking place, but we do not believe it can be
responsible for the size of the observed difference between the
\citet{Beswick08} results and the other data.  It would also not
explain the difference between the essentially constant value of
$q_{24}$ found in this work and by \citet{Boyle07}, and the variation
with 24-$\umu$m flux density found by \citet{Beswick08}.

The \citet{Beswick08} sample is the only stacking study which is
working with resolved radio sources -- however, the flux density
measurement method that they use should include all of the resolved
flux density.  Any loss of radio flux would lead to an increase in
$q_{24}$, rather than the lower value that they find compared with
other works.

None of the explanations for the discrepancy in stacking results are
satisfactory -- while we believe that the most likely explanation is
some combination of an effect due to the differing noise levels and a
stacking bias, this remains speculation.  Nonetheless, all the
indications are that the same effect is occurring equally to all radio
sources in a field, and so while the absolute values of $q_{24}$ and
$q'_{24}$ are uncertain, their dependence on infrared flux density
will not be affected.  We find no evidence for any variation in the
value of $q_{24}$ or $q'_{24}$ over the 24-$\umu$m flux density range
of 150-$\umu$Jy and 10~mJy, and no evidence for a variation in the
value of $q'_{70}$ over the 70-$\umu$m flux density range 10 -- 100~mJy.

\section{Conclusions}
We have compared the two methods commonly used in stacking
experiments, and demonstrated that they give comparable measurements
of the average radio flux density for binned sources.  We have shown
that the median flux density is a much more reliable estimator than
the mean, and that the noise level of stacked images decreases as
$1/\sqrt{N}$, at least until a depth of $\sim0.5$~$\umu$Jy~beam$^{-1}$
at 1.4~GHz.  Creating stacked images leads to less information being
retained on the flux density distribution of sources, and we argue
that future stacking experiments should not make images, but instead
work directly from the individual flux density measurements that can
be calculated at each source position.

We have calculated $q_{24}$ and $q'_{24}$ from sources within the xFLS
field, and demonstrated that they can be related to each other with
the simple assumption that all radio sources have the spectral index
of $\alpha=0.4$, over the 24-$\umu$m flux density range of 150~$\umu$Jy\
-- 10~mJy.  Using this conversion, we have compared the stacking
results of \citet{Boyle07} to the results obtained from our three
SWIRE field images, and demonstrated that they are consistent, within
the errors that can be placed on the median stacked values of
$q_{24}$.  There appears to be some field-to-field variation seen
between the xFLS and SWIRE fields, with stacking results from the xFLS
field leading to a value of $q_{24}$ that is $\sim0.3$ lower than the
value seen in the SWIRE fields.  This variation is also seen when
stacking radio sources using the 70-$\umu$m catalogues of each field.
Through comparison with the results of \citet{Appleton04}, who look at
individual sources which are detected in the xFLS field, we conclude
that our radio flux density measurements are not being significantly
biased by the stacking procedure.

We have considered several potential explanations for the difference
in the median values of $q_{24}$, and shown that effects such as a
flux calibration error cannot be responsible for the
difference in results.  While we can not explain the systematic offset
which is being seen, we believe that it originates in some combination
of a `stacking bias' and the varying noise levels of the different
radio surveys.  The systematic effect leads to an uncertainty in the
absolute values of $q_{24}$ and $q_{70}$, but not in their variation
with infrared flux density.

We find no evidence for a variation in the median value of $q_{24}$
from any of the survey fields down to a 24-$\umu$m flux density of
150~$\umu$Jy.  This is in agreement with the results of
\citet{Boyle07}, although is in contrast to the tentative findings of
\citet{Beswick08}.  We find a similar result at 70~$\umu$m, although
over the brighter flux density range of 10 -- 100~mJy.  We conclude
that there is no evidence for any significant variation in the
infrared / radio correlation with infrared flux density, down to the
limits set by the extremely sensitive {\it Spitzer} observations of
the xFLS and SWIRE fields.

\section*{Acknowledgments}
TG thanks the UK STFC for a Studentship.  We thank the staff of the
GMRT who have made these observations possible.  The GMRT is operated
by the National Centre for Radio Astrophysics of the Tata Institute of
Fundamental Research, India.  

\bibliography{./References}
\bibliographystyle{./mn2e}
\label{lastpage}

\end{document}